\documentclass[amsmath,amssymb,aps,prc,twocolumn,superscriptaddress,showpacs,preprintnumbers]{revtex4}

\usepackage{graphicx}
\usepackage{multirow}
\usepackage{times}
\usepackage{color}
\usepackage[normalem]{ulem}  
\renewcommand\sout{\bgroup \color{red} \ULdepth=-.5ex \ULset}
\newcommand{\tilA}{\tilde{A}}
\newcommand{\Kp}{K^\prime}
\newcommand{\cc}{\text{c.c.}}
\newcommand{\half}{\frac{1}{2}}

\newcommand{\delt}{\partial_t}
\DeclareMathOperator{\tr}{tr}
\newcommand{\cn}{\mathrm{cn}\,}
\newcommand{\diag}{\mathrm{diag}\,}
\newcommand{\vf}{\boldsymbol f}
\newcommand{\va}{\boldsymbol a}

\newcommand{\vphi}{\boldsymbol \phi }

\newcommand{\rp}{\right)}
\newcommand{\lp}{\left(}
\newcommand{\rb}{\right\}}
\newcommand{\lb}{\left\{}
\newcommand{\rbb}{\right]}
\newcommand{\lbb}{\left[}
\newcommand{\pT}{p_{\scriptscriptstyle{T}}}
\newcommand{\PSfig}[2]{\includegraphics[width=#1]{#2}}
\begin{document}
\preprint{KUNS-2529, YITP-14-89}

\title{
Parametric Instability of Classical Yang-Mills Fields in a Color Magnetic Background
}

\author{Shoichiro Tsutsui}
\email[]{tsutsui@yukawa.kyoto-u.ac.jp}
\affiliation{Department of Physics, Faculty of Science, Kyoto University,
Kyoto 606-8502, Japan}
\affiliation{Yukawa Institute for Theoretical Physics, Kyoto University,
Kyoto 606-8502, Japan}
\author{Hideaki Iida}
\affiliation{Department of Physics, Faculty of Science, Kyoto University,
Kyoto 606-8502, Japan}
\author{Teiji Kunihiro}
\affiliation{Department of Physics, Faculty of Science, Kyoto University,
Kyoto 606-8502, Japan}
\author{Akira Ohnishi}
\affiliation{Yukawa Institute for Theoretical Physics, Kyoto University,
Kyoto 606-8502, Japan}

\date{\today}

\begin{abstract}
We investigate instabilities of classical Yang-Mills fields in a time-dependent spatially homogeneous color magnetic background field in a non-expanding geometry for elucidating the earliest stage dynamics of ultra-relativistic heavy-ion collisions.
The background gauge field configuration considered in this article is spatially homogeneous and temporally periodic, and is introduced by Berges-Scheffler-Schlichting-Sexty (BSSS).
We discuss the whole structure of instability bands of fluctuations around the BSSS background gauge field on the basis of Floquet theory,
which enables us to discuss the stability in a systematic way.
We find various instability bands on the $(p_z, p_T)$-plane.
These instability bands are caused by parametric resonance despite the fact that the momentum dependence of the growth rate for $|\mathbf{p}| \leq \sqrt{B}$ is similar to a Nielsen-Olesen instability.
Moreover,
some of instability bands are found to emerge not only in the low momentum but also in the high momentum region; 
typically of the order of the saturation momentum as $|\mathbf{p}| \sim \sqrt{B} \sim Q_{\rm s}$.
\end{abstract}

\maketitle

\section{Introduction}
Remarkable properties of the quark-gluon plasma (QGP) have been revealed by the recent ultra-relativistic heavy-ion experiments at the Relativistic Heavy-Ion Collider (RHIC) at Brookhaven National Laboratory and the Large Hadron Collider (LHC) at CERN.
Hydrodynamic models turned out to be successful in describing the transverse momentum ($\pT$)-spectra and the anisotropic flows ($v_n$) of hadrons~\cite{Heinz05,Teaney09,Hirano12}.
The observation of large elliptic flow parametrized by $v_2$ suggests two important features of QGP;
nearly perfect fluidity and early thermalization.
The initial spatial eccentricity of the participants seems to be efficiently converted to the final momentum anisotropy.
This is only possible when the viscosity is small enough and the pressure is developed in the early stage.
Hydrodynamic phenomenology suggests that shear viscosity of QGP is $\eta/s = (1-3)/4\pi$.
Hydrodynamic analyses also require a short thermalization time, $\tau_\mathrm{th} = (0.6-1.0) \mathrm{fm}/c$ which is significantly shorter than that evaluated from transport theories~\cite{Baier01,Baier02}.
There are no conclusive scenarios found yet to explain thermalization in the far-from-equilibrium stage of heavy-ion collisions.

Some of the promising mechanisms for early thermalization are instabilities which cause rapid growth of a classical Yang-Mills (CYM) field followed by its decay into particles.
CYM field theory is believed to be a good starting point for describing the earliest stage of heavy-ion collisions.
In the high energy limit,
nuclear wave functions are well expressed by color glass condensate (CGC) effective field theory~\cite{Gelis10a,Gelis12a}.
In the framework of CGC,
the classical solution gives transversely polarized color electromagnetic fields whose sources are valence partons in the large $x$-region.
The contact of two nuclei converts CGC into the state with longitudinally polarized color electromagnetic fields called glasma~\cite{Lappi06}.
Classical fields in glasma show instabilities, and some of classical gluon fields grow exponentially, show chaoticity and may decay into particles via field-particle conversions.
Thus instabilities of classical fields should play important roles in thermalization in heavy-ion collisions~\cite{Arnold05,Romatschke06a,Romatschke06b,Berges08,Berges09,Kunihiro08,Kunihiro10,Iida13,Epelbaum14}.

It is known for a long time that an instability occurs in electromagnetic plasmas when anisotropy is present.
When the particle momentum distribution is anisotropic, the particle current and the background magnetic field enhance each other. 
This is called the Weibel instability~\cite{Weibel59}.
The Weibel instability of the color magnetic field is also expected to emerge in glasma, and has been discussed as one of the triggers leading to early thermalization in heavy-ion collisions~\cite{Mrowczynski88,Arnold03,Romatschke06a,Romatschke06b}.
The system under a homogeneous and static color magnetic field shows a different instability.
Under a homogeneous color magnetic field,
the spin-magnetic field interaction makes the lowest Landau level negative for the spin one system.
If this is the case, the resultant instability called the Nielsen-Olesen instability~\cite{Nielsen78}, is also expected as a triggering mechanism of the early thermalization in heavy-ion collisions ~\cite{Fujii08,Fujii09,Iwazaki09}.

This is not the end of the story.
Yet another instability can occur under a homogeneous but time dependent color magnetic field. 
This type of instability is alluded by Berges, Scheffler, Schlichting and Sexty (BSSS)~\cite{Berges12a}.
Their analysis based on the classical statistical simulation suggests that low momentum modes become unstable under the time-dependent color magnetic field.
This instability is seemingly reminiscent of the Nielsen-Olesen instability, because it is caused by the homogeneous color magnetic field and the dominant growth rate has similar longitudinal momentum dependence to that of the Nielsen-Olesen instability.
They also suggest that there exists a sub-dominant instability band in a high momentum region.
It is caused by the time dependence of the background field, and thus the underlying nature of the sub-dominant instability is thought to be induced by parametric resonance.
The Nielsen-Olesen instability and the parametric-resonance-induced instability
seem to coexist in their study.

However, the analysis on the nature of the instability has some ambiguous points to be further elucidated.
The Nielsen-Olesen gauge configuration and BSSS gauge configuration are not connected with each other by any gauge transformations.
Moreover, gauge fluctuations do not form Landau orbit under the BSSS configuration, since not only the color magnetic field but also the background gauge field is homogeneous.
Then one may wonder what is the genuine nature of the instability induced by the homogeneous but time-dependent color magnetic fields; parametric or Nielsen-Olesen instability? Or does the one induce the other?

In this article, 
we perform a systematic investigation of the instabilities of classical gluon fields under the homogeneous but time-dependent background color magnetic fields
in the linear regime.
Specifically, we consider the BSSS initial condition~\cite{Berges12a} shown in Eq.~\eqref{nonabelconf} for the background field, whose solution is known to be the Jacobi elliptic function.
This setup may not be very realistic for heavy-ion collisions but highly idealized one, where the gauge configuration is rapidity-independent, no color flux tubes, absence of the longitudinal color electric fields.
But at the same time, the configuration is similar to the glasma because the resultant color magnetic field is originated from the non-abelian nature of QCD.
Therefore, it should provide insight into the realistic situation by studying the time evolution from the BSSS initial condition as noted in Ref~\cite{Berges12a}.
We analyze the stability of fluctuations around the BSSS background gauge field systematically on the basis of the Floquet theory, which consists the basis of the Bloch theory.
In this setup, we can precisely obtain growth rates of the fluctuations by solving the equations of motion for a given momentum during one period of the background field and by evaluating the eigenvalues of a $3(N_c^2-1) \times 3(N_c^2-1)$ matrix called a monodromy matrix for color SU($N_c$).
As a result, we get the complete structure of the instability bands in the whole momentum region not only in the longitudinal but also transverse directions caused by parametric resonance.

Parametric resonance plays an important role in many fields of physics.
For instance, it would contribute to preheating in the early universe in cosmic inflation~\cite{Kofman:1997yn,Kaiser:1997hg,Greene:1997fu,Finkel:2000nu}.
Parametric resonance also might trigger thermalization in heavy-ion collisions because it can give rise to rapid particle production.
Instability due to parametric resonance in O($N$) scalar field theory has been analyzed and an exponential growth of the particle number is demonstrated in numerical analyses~\cite{Berges02c,Berges05a,Berges12b}.
The present analysis should give a general and lucid mathematical basis of the parametric resonance or {\em parametric instability} ubiquitous in many fields of physics.

This article is organized as follows.
In Sec.~\ref{sec:instability},
we explain our setup, fluctuations of the CYM field around a homogeneous time-dependent color magnetic field.
We also give basics of parametric instability and a brief overview of the Floquet theory, which is applied to analyze instability bands.
In Sec.~\ref{sec:band}, 
we show numerical results of instability bands of Yang-Mills fields.
Finally, 
we give a summary and discuss the relevance of these instabilities to the thermalization in the early stage of heavy-ion collisions in Sec.~\ref{sec:summary}.

\section{Instabilities under a strong color magnetic field}\label{sec:instability}
We discuss instabilities of fluctuations of the CYM field under the BSSS background field.
In Sec.~\ref{subsec:EOM}, we derive the linearized equation of motion (EOM) of the fluctuations and show that the EOM is a special case of a Hill\rq{}s differential equation.
It is well known that solutions of a Hill\rq{}s differential equation show instabilities called parametric resonance.
In Sec.~\ref{subsec:PI}, we review basics of parametric resonance according to concrete examples.
In Sec.~\ref{sec:Floquet}, we give a brief overview of the Floquet theory which is a general mathematical framework to determine instability bands for a given equation with a periodic coefficient. 
The reader already familiar with these topics can skip to Sec.~\ref{app to YM} where we apply the Floquet theory to CYM theory.

\subsection{CYM equation under a homogeneous color magnetic field}\label{subsec:EOM}
We briefly summarize CYM equations for the background field and fluctuations under a homogeneous color magnetic field.
Throughout this article,
we take the temporal gauge $A^a_0 = 0$ with a homogeneous background color magnetic field in a non-expanding geometry.

In SU($N$) pure Yang-Mills theory, color magnetic fields are defined by
\begin{align}
B^a_i 
= 
\epsilon_{ijk} \lp \partial_j A^a_k - \frac{1}{2}f^{abc}A^b_j A^c_k \rp,
\label{defmag}
\end{align}
where $A_i^a$ is a gauge field and $f^{abc}$ is the structure constant.
The superscripts $a,b,\dots$ and the subscripts $i,j,\dots$ denote color and Lorentz indices, respectively.
The gauge coupling constant is included in the definition of the gauge fields.

There are two types of gauge configurations to make homogeneous color magnetic fields.
One is the abelian configuration such as 
\begin{align}
A^3_x = -\half By , \quad A^3_y = \half Bx .
\label{abelconf}
\end{align}
The Nielsen-Olesen instability is induced by the color magnetic field in the above configuration.
Because of the spatial dependence of the background gauge field, the transverse motion of gluons is quantized to form Landau levels.
The Nielsen-Olesen instability is caused by the particles in the lowest Landau level, whose eigenfrequency becomes complex due to the spin-magnetic field interaction.

The other configuration is the non-abelian configuration given as
\begin{align}
A^a_i = \tilA(t) \lp \delta^{a2}\delta_{ix} + \delta^{a1}\delta_{iy} \rp ,
\label{nonabelconf}
\end{align}
which depends on time but not on spatial coordinates.
Therefore, both color magnetic and gauge fields are homogeneous.
As a consequence, 
$p_x$ and $p_y$ are  good quantum numbers.	 
This point is completely different from the former case.
Note that there also exist homogeneous color electric fields $E^2_x$ and $E^1_y$, since color electric fields are defined by $E^a_i = \dot{A}^a_i$ in the temporal gauge.
	
The instability of a few low momentum modes under the non-abelian configuration was first discussed by BSSS~\cite{Berges12a}.
The classical Yang-Mills equation is given by
\begin{align}
\ddot{A}_i^a - (D_j F_{ji})^a = 0 ,
\label{CYM}
\end{align}
which is reduced to
\begin{align}
\ddot{\tilA} + \tilA^3 = 0 ,
\end{align}
by virtue of Eq.~\eqref{nonabelconf}.

The solution for the background field is given by the Jacobi elliptic function $\cn(t;k)$ since
it satisfies the following equation,
\begin{align}
	y^{\prime\prime} + (1-2k^2)y + 2k^2y^3= 0 .
\end{align}
Note that we use modulus $k$ as the second argument of $\cn(t;k)$\cite{Whittaker27}, which is different from the notation in Ref~\cite{Berges12a}.

For instance, with the initial condition as $\tilA(t=0)=\sqrt{B_0}$ and $\dot{\tilA}(t=0)=0$, the solution of this equation is given by
\begin{align}
\tilA(t) = \sqrt{B_0} \cn{ \lp \sqrt{B_0}t;1/\sqrt{2} \rp }.
\label{tilA}
\end{align}
The period of $\tilA$ is given by the complete elliptic integral of the first kind $K(k)$;
$T=4K(1/\sqrt{2})/\sqrt{B_0} \simeq 7.42/\sqrt{B_0}$.
In this way, the background gauge field is a periodic function in time.

By the shift $A^a_i \rightarrow A^a_i + a^a_i$, we get the EOM of the fluctuations described by $a^a_i$.
Since the background gauge field is homogeneous, we can work with the EOM for each Fourier component of fluctuations in the linear regime. 
The linearized EOM for $a^a_i$ is given by
\begin{align}
\ddot{a}^a_i = - {\Omega^2[\tilA(t)]}^{ab}_{ij} a^b_j ,
\label{Hill}
\end{align}
where ${\Omega^2[\tilA]}^{ab}_{ij}$ is a $9\times9$ matrix as
\begin{align}
{\Omega^2[\tilA]}^{ab}_{ij}
&=
(-D_kD_k\delta_{ij} + D_iD_j + 2iF_{ij})^{ab} \\
&=
(p^2\delta_{ij} - p_ip_j) \delta^{ab}\notag \\
&+ i\tilA ( -2p_x\delta_{ij} + p_i\delta_{jx} + p_j\delta_{ix} ) f^{a2b}\notag \\ 
&+ i\tilA ( -2p_y\delta_{ij} + p_i\delta_{jy} + p_j\delta_{iy} ) f^{a1b}\notag \\
&-\tilA^2 \delta_{ij} ( f^{a2d}f^{d2b} + f^{a1d}f^{d1b} ) \notag \\
&+\tilA^2 ( f^{a2d}\delta_{ix} + f^{a1d}\delta_{iy} )( f^{d2b}\delta_{jx} + f^{d1b}\delta_{jy} ) \notag \\
&+2\tilA^2 f^{a3b} (\delta_{ix}\delta_{jy} - \delta_{iy}\delta_{jx}) .
\label{Omega}
\end{align}
$f^{abc}$ is the structure constant of SU($N$).
When the background field is given by Eq.~\eqref{nonabelconf}, SU(2) components, namely, $A^1_i$, $A^2_i$ and $A^3_i$ are decoupled from other components. Hereafter, we concentrate on SU(2) Yang-Mills theory, where the the structure constant is given by $\epsilon^{abc}$.

Without loss of generality,
we can take $p_y=0$ due to the rotational symmetry in the transverse direction.
Then, we can easily find that the coefficient matrix $\Omega^2$ is a block diagonalized matrix as
\begin{align}
\Omega^2 = \diag(\Omega^2_4,\Omega^2_5) ,
\end{align}
where $\Omega^2_4$ and $\Omega^2_5$ are $4\times 4$ and $5\times 5$ matrices, respectively.
The explicit forms of the matrices are given in Appendix~\ref{sec:Hill}.
Thus, the linearized EOM for $a^a_i$ is decomposed into two sectors;
\begin{align}
\ddot{a}_\alpha &= - {\Omega^2_4[\tilA(t)]}_{\alpha\beta} a_\beta ,
\label{Hill4} \\
\ddot{a}_A &= - {\Omega^2_5[\tilA(t)]}_{AB} a_B .
\label{Hill5}
\end{align}
We use following notation; $\alpha, \beta, \dots = (1y,2x,2z,3y)$ and $A,B,\dots = (1x,1z,2y,3x,3z)$.

Equation~\eqref{Hill} or Eqs.~\eqref{Hill4} and~\eqref{Hill5} are second order linear ordinary differential equations with a periodic coefficient matrix.
This type of ordinary differential equation is called Hill\rq{}s equation.
In general, Hill\rq{}s equation has unstable solutions due to the periodicity of the coefficient matrix~\cite{Magnus}.

It should be noted that the physical solutions must satisfy the Gauss's law.
The background field Eq.~\eqref{nonabelconf} satisfies $D_i\dot{A_i^a}=0$. 
After shifting $A^a_i \rightarrow A^a_i + a^a_i$ and picking up $\mathcal{O}(a)$ terms from the
Gauss's law, $D_i (\dot{A_i} + \dot{a_i})^a = 0$, we find 
\begin{align}
ip_i \dot{a}^a_i + \epsilon^{abc} 
\lbb 
\delta^{b2} \lp \tilA\dot{a}^c_x - \dot{\tilA}a^c_x \rp 
+ \delta^{b1}  \lp \tilA\dot{a}^c_y - \dot{\tilA}a^c_y \rp
\rbb
=0 ,
\label{Gauss}
\end{align}
which is to be imposed on  $a^a_i$ at the initial time.

\subsection{Parametric instability}\label{subsec:PI}
In this subsection,
we give a brief account of general aspects of instabilities under a periodic perturbation.
These phenomena are well known as parametric resonances or parametric instabilities.
First, we consider Mathieu\rq{}s equation in order to see how parametric instabilities occur. 
Mathieu\rq{}s equation is one of the simplest Hill\rq{}s equation which has non-trivial instabilities;
\begin{align}
\ddot{f} = - \lp \lambda + 2\epsilon \cos t \rp f .
\label{Mathieu}
\end{align}
For simplicity, we assume $\lambda>0$ and $\epsilon \ll 1$. 
We can investigate the stability of the solutions of Eq.~\eqref{Mathieu} in a perturbative way when the external force term, $2\epsilon \cos t$, can be regarded as small.
When we expand $f(t)$ as $f = f_0 + \epsilon f_1 + \dots$, the lowest order solution is given by $f_0 = A_0 e^{i\sqrt{\lambda}t} + \text{c.c}$, and $f_1$ follows
\begin{align}
\ddot{f_1} + \lambda f_1 
= 
- A_0 \lp e^{i(\sqrt{\lambda}+1)t} + e^{i(\sqrt{\lambda}-1)t} \rp + \text{c.c.~} .
\end{align}
This is the EOM for a driven oscillator.
Its eigenfrequency is $\sqrt{\lambda}$ and the frequencies of external forces are $\sqrt{\lambda}\pm1$.
If $\lambda = 1/4$, the oscillator resonates and becomes to be amplified.
In general, such resonance occurs if $\lambda = n^2/4 \, (n=1,2,\dots)$.
Moreover, there are more sophisticated perturbative techniques to determine instability boundaries $\lambda = \lambda(\epsilon)$~\cite{Bender78}.
For example, the instability boundaries of Mathieu\rq{}s equation passing through the point $(\lambda, \epsilon)=(1/4,0)$ are given by 
$\lambda =1/4 \pm \epsilon - \epsilon^2/2 + \mathcal{O}(\epsilon^3)$.

For the purpose to analyze instabilities of CYM fields,
it is instructive to consider Lam\'e\rq{}s equation.
Lam\'e\rq{}s equation is a little more complicated than Mathieu\rq{}s equation, which has the elliptic function as an external force term instead of $\cos t$;
\begin{align}
\ddot{f} = - \lp \lambda + \epsilon\cn^2(t;k) \rp f .
\label{Lame}
\end{align}
In fact, we will see that Lam\'e\rq{}s equations with $\epsilon = \pm1, 3$ and $k=1/\sqrt{2}$ are obtained for some momentum modes in the linearized EOM for fluctuations, Eqs.~\eqref{Hill4} and~\eqref{Hill5}.
In particular, Lam\'e\rq{}s equation with $\epsilon = -1$ leads to the largest growth rate of the above three cases, and it also describes the CYM equation for the fluctuation mode having the maximum growth rate.
We also mention that these equations have a good property in an analytical point of view.
Lam\'e\rq{}s equations with $\epsilon = 1, 3$ and $k=1/\sqrt{2}$ are exactly solvable to get closed form solutions~\cite{Greene:1997fu}. 
When $\epsilon = -1$, any closed form solution is not known, but its solution has been investigated analytically~\cite{Berges12a}.
The perturbative approach mentioned above is also a general framework and it can be performed in a parallel way as the analysis for Mathieu\rq{}s equation, but its applicability is still limited.
It is valid only for $0 \leq \epsilon <1$.
The details of perturbative calculations are presented in Appendix~\ref{App:Lame}.

Instead of these analytical techniques, we will use a more general framework to find unstable modes together with their growth rates utilizing numerical calculations.

\subsection{Floquet theory}\label{sec:Floquet}
We can perform precise stability analyses of the linearized EOMs,
Eqs.~\eqref{Hill4} and~\eqref{Hill5}, by using the Floquet theory, even though it is difficult to obtain analytic solutions.
	
In this subsection, we give a brief overview of the Floquet theory
(see also the appendix of \cite{Dusling:2010rm}).

Suppose an ordinary differential equation of order $n$ have a $T$-periodic coefficient $P(t)$, i.e.
\begin{align}
\frac{d\vf}{dt} = P(t)\vf,
\quad P(t+T) = P(t) ,
\label{ODE}
\end{align}
where $\vf$ is a $n$-dimensional vector and $P(t)$ is an $n\times n$ matrix.
The fundamental matrix of Eq.~\eqref{ODE} is defined by $n$ independent solutions $\{\vphi_i\}_{i=1,\dots,n}$;
\begin{align}
\Phi(t) = \lp \vphi_1(t) ,\dots , \vphi_n(t) \rp .
\end{align}
If $\Phi(t)$ is a fundamental matrix, $\Phi(t+T)$ is also a fundamental matrix
due to the periodicity of the coefficient matrix, $P(t)$.
Then, there exists a constant matrix $M$ such that $\Phi(t+T) = \Phi(t)M$.
$M$ is called a monodromy matrix.
By construction, $\Phi(t)$ is a regular matrix and we can get the monodromy matrix $M$ as
\begin{align}
M = \Phi(0)^{-1} \Phi(T) .
\label{monodromy}
\end{align}
It should be noted that $\det M \neq 0$ since $\Phi$ is regular.
Specifically, it holds that $\det M = 1$ in the case $\tr P = 0$, which is fulfilled in Hamiltonian systems.
This fact follows from the Liouville's theorem since $\det M$ is equivalent to the Jacobian of phase space variables at $t=0$ and $t=T$.
The fundamental matrix is represented by the monodromy matrix as
\begin{align}
\Phi(t) = F(t)\exp\lp (\log M)\frac{t}{T}\rp ,
\label{Floquet thm}
\end{align}
where $F(t)$ is a $T$-periodic matrix.
The specific form of $F(t)$ is not relevant for our discussion.
The eigenvalues of $M$ are called characteristic multipliers and we denote them as $\mu_1,\dots,\mu_n$.
Characteristic multipliers determine the long-time behaver of the solutions.
We can categorize the stability at $t>0$ by using the characteristic multipliers as follows.
\begin{enumerate}
\item If $|\mu| > 1$, the solution exponentially diverges. 
\item If $|\mu| = 1$, the solution is (anti)periodic or polynomially diverges. 
\item If $|\mu| < 1$, the solution is bounded. 
\end{enumerate}
Thus, when $|\mu| > 1$, the solution is unstable and the growth rate of the unstable solution is given by the exponent $(\log\mu)/T$ according to Eq.~\eqref{Floquet thm}.
This exponent is sometimes called a characteristic exponent.
When $|\mu| = 1$, the solution can be also unstable with a polynomial growth.
This is caused by the degeneracy of $M$.
In Hamiltonian systems, the maximum multiplier $|\mu|_\mathrm{max}$ is bounded as $|\mu|_\mathrm{max}\geq1$ since $\det M = 1$.

Let us take a single component Hill\rq{}s equation;
\begin{align}
\ddot f = - \omega^2(t)f ,
\end{align}
where $\omega^2(t+T)=\omega^2(t)$.
In general,
a second order equation can be transformed into a first-order equation with two components. 
In fact, by putting $\vf = (f, \dot{f})$, we have
\begin{align}
\frac{d\vf}{dt}
=
\begin{pmatrix}
0 & 1 \\ -\omega^2(t) & 0
\end{pmatrix}
\vf.
\label{single Hill}
\end{align}
This equation can be easily analyzed as we will see.

Because of the conservation of the Wronskian, 
$\det\Phi(t)=f_1 \dot{f}_2-\dot{f}_1 f_2$, we find $\det M=\mu_1\mu_2=1$ by taking the determinant of Eq.~\eqref{monodromy}.
Therefore, the eigenvalues of the monodromy matrix follow the characteristic equation;
$\mu^2 - (\tr M) \mu +1 =0$.
Thus, the stability of the solution of a single component Hill\rq{}s equation is governed by $\tr M$ as
\begin{enumerate}
\item If $|\tr M| > 2$, the solution exponentially diverges. 
\item If $|\tr M| = 2$, the solution is (anti)periodic or linearly diverges. 
\item If $|\tr M| < 2$, the solution is bounded. 
\end{enumerate}
This fact simplifies the stability analysis of a single component Hill\rq{}s equation because we can discuss the stability without diagonalizing $M$.

To see how the general Floquet analysis is performed, we shall go back to the stability analysis of the Lam\'e\rq{}s equation~\eqref{Lame}, where the periodic coefficient $\omega^2(t)$ is given by the elliptic function.
Here, the parameters $\lambda$ and $\epsilon$ are arbitrary.
As we mentioned in the previous subsection, the linearized CYM equations, Eqs.~\eqref{Hill4} and~\eqref{Hill5}, take a form of Lam\'e\rq{}s equation for some momentum modes.
For instance,
Eq.~\eqref{Hill4} becomes block-diagonalized and is decomposed into two simultaneous equations for zero transverse momentum modes ($p_x=p_y=0$) or zero longitudinal momentum modes ($p_z=0$).
The EOM for these modes in CYM theory are summarized in Appendix~\ref{sec:Hill}.

We here consider the equation for $(a^1_y, a^2_x)$ of zero transverse momentum ($p_x=p_y=0$) shown in Eq.~\eqref{systemB}.
We can decompose the EOM Eq.~\eqref{systemB} into two equations which give Lam\'e\rq{}s equation with $\epsilon = -1,3$.
We also consider an equation for $a_{2z}$ of zero longitudinal momentum ($p_z=0$) shown in Eq.~\eqref{systemG}.
This gives Lam\'e\rq{}s equation with $\epsilon = 1$;
\begin{align}
a^{\prime\prime}_+(\theta;p_z,p_T=0) 
&= -  \lp p_z^2/B_0 + 3\cn^2(\theta) \rp a_+ 
\label{Lame3}, \\
a^{\prime\prime}_-(\theta;p_z,p_T=0) 
&= -  \lp p_z^2/B_0 - \cn^2(\theta) \rp a_- 
\label{Lame-1}, \\
a^{\prime\prime}_{2z}(\theta;p_z=0,p_T) 
&= -  \lp p_T^2/B_0 + \cn^2(\theta) \rp a_{2z} 
\label{Lame1},
\end{align}
where $a_\pm = a^1_y \pm a^2_x$ and $\theta = \sqrt{B_0} t$.
Primes denote derivatives with respect to $\theta$.
Unless otherwise noted, the modulus of elliptic function is $k=1/\sqrt{2}$.
We note that  Floquet analysis can be done by a quite simple numerical calculations.
In addition, there is no constraint for $\epsilon$ to apply Floquet theory to Lam\'e\rq{}s equation.

\begin{figure}[bth]
\PSfig{8.5cm}{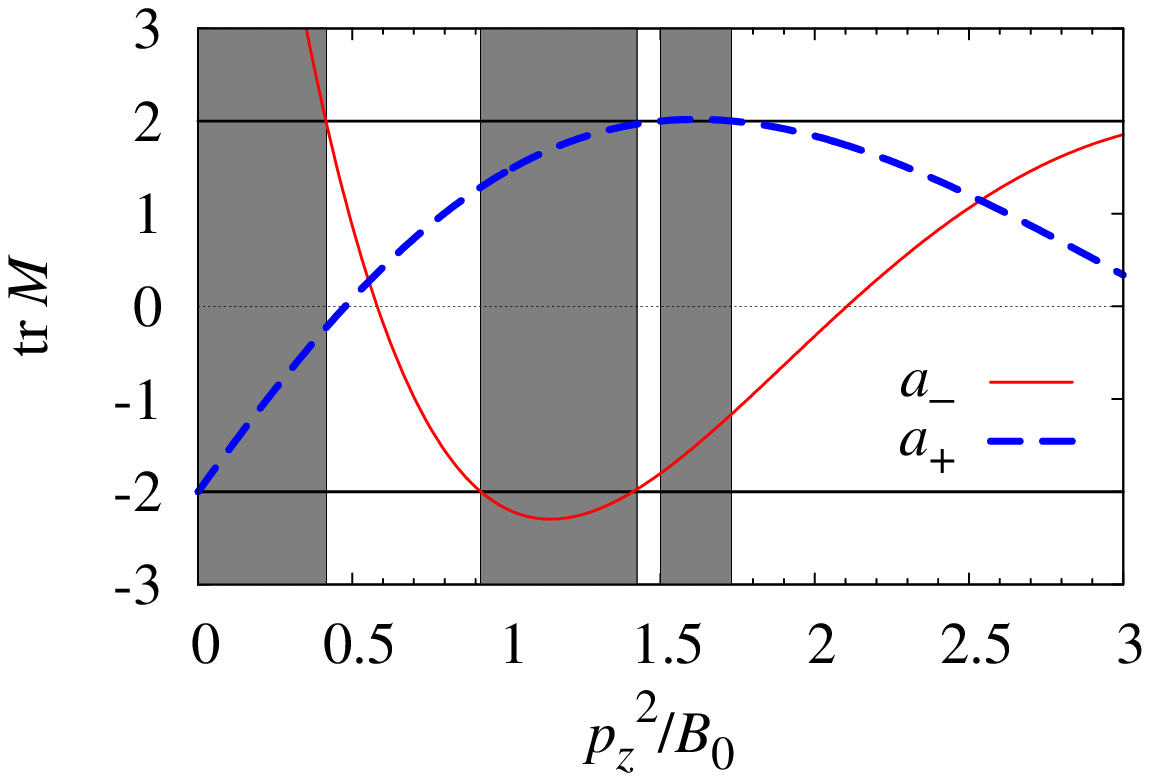}
\PSfig{8.5cm}{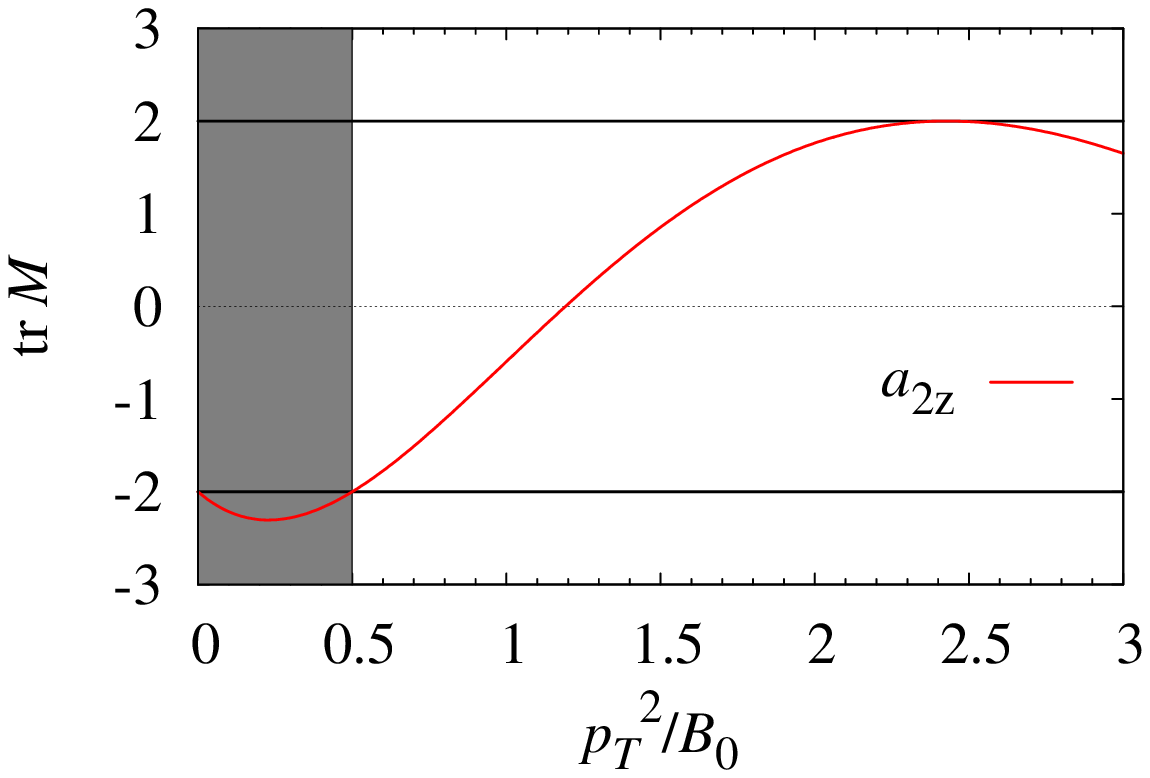}
\caption{
Floquet analysis for Eqs.~\eqref{Lame3}, \eqref{Lame-1} and \eqref{Lame1}.
The shaded area denotes the instability bands specified by the $p_z$ or $p_T$-region with $|\tr M| \geq 2$.
These are the consequence of parametric resonance.
The instability bands of $a_-$ (Lam\'e\rq{}s eq. with $\epsilon =  -1$) are $0 \leq p_z^2/B_0 \leq 0.41$ and $0.91 \leq p_z^2/B_0 \leq 1.42$.
The instability bands of $a_+$ (Lam\'e\rq{}s eq. with $\epsilon =  3$) is $3/2 \leq p_z^2/B_0 \leq \sqrt{3}$.
The instability bands of $a_{2z}$ (Lam\'e\rq{}s eq. with $\epsilon =  1$) is $0 \leq p_T^2/B_0 \leq 1/2$.
Note that exact instability boundaries are known for $\epsilon =  1,3$~\cite{Greene:1997fu}.
Our numerical calculation agrees with that.
}
\label{Lames}
\end{figure}

Figure~\ref{Lames} shows $\tr M$ of Eqs.~\eqref{Lame3},~\eqref{Lame-1} or~\eqref{Lame1} as a function of $p_z^2/B_0$ or $p_T^2/B_0$.
The instability bands are specified by $|\tr M| \geq 2$.
In this calculation,
we set the initial fundamental matrix as a unit matrix, i.e. $\Phi(t=0)=1$.
All we have to do is to solve Eqs.~\eqref{Lame3} and \eqref{Lame-1} for a period $T$ numerically at a given momentum.
After solving them, we easily get $\tr M = \tr\Phi(T)$ by Eq.~\eqref{monodromy} as a function of squared momentum normalized by the initial background magnetic field.

The emergence of the first instability band of $a_-$ ($0 \leq p_z^2/B_0 \leq 0.41$) is easily expected from the form of Eq.~\eqref{Lame-1}.
The eigenfrequency of $a_-$ become complex for sufficiently small momentum $p_z^2 \ll B_0$ since it is approximately given by $\omega \sim \sqrt{p_z^2 - B(t)}$.
On the other hand,
the instability boundary is modified from the naively expected one $p_z^2/B_0 = 1$ to $p_z^2/B_0 = 0.41$ due to time dependence of the background field. 
The remarkable feature of this band is that the unstable modes in this region has quite large growth rate since $\tr M \gg 2$.
The same band structure appears in Yang-Mills theory, as we will see later.

We also find the instability band of $a_+$ ($3/2 \leq p_z^2/B_0 \leq \sqrt{3}$), the second band of $a_-$ ($0.91 \leq p_z^2/B_0 \leq 1.42$) and the instability band of $a_{2z}$ ($0 \leq p_T^2/B_0 \leq 1/2$).
These instabilities are the consequence of parametric resonance, and are not intuitively expected from the forms of Eqs.~\eqref{Lame3},~\eqref{Lame-1} and~\eqref{Lame1}.

\subsection{Application to Yang-Mills theory}\label{app to YM}
The Floquet analysis of a multi-component Hill\rq{}s equation can be done in much the same way as the single-component analysis.
In this subsection, we apply the Floquet theory to the CYM equation.
Equation (\ref{Hill}) is also transformed into the first order equation by introducing color electric fields $e^a_i = \dot{a}^a_i$.
\begin{align}
\frac{d}{dt}
\begin{pmatrix}
a^a_i \\ e^a_i
\end{pmatrix}
=
\begin{pmatrix}
0 & \delta^{ab}\delta_{ij} \\ -[\Omega^2(t)]^{ab}_{ij} & 0
\end{pmatrix}
\begin{pmatrix}
a^b_j \\ e^b_j
\end{pmatrix}.
\label{Eq:Hill_1st}
\end{align}
The simplest way to calculate characteristic multipliers is to set the initial fundamental matrix as a unit matrix as we have done in the previous subsection.
The solutions starting from the unit matrix initial condition
constitute a complete set,
and any solution of the equation of motion Eq.~\eqref{Eq:Hill_1st}
is represented by a linear combination of these solutions.

For the Yang-Mills field, the situation is somewhat complicated because the initial condition $\Phi(0) = 1$ does not satisfy the Gauss\rq{}s law (\ref{Gauss}).
As a consequence, 
we need to worry about picking up the instabilities of unphysical channel.
We shall show here how to extract physical instability bands in the framework of the Floquet theory.
At $t=0$, Gauss\rq{}s law reads
\begin{align}
ie^a_ip_i/\sqrt{B_0}  + \epsilon^{abc} 
\lp
\delta^{b2} e^c_x 
+ \delta^{b1} e^c_y
\rp
=0 .
\label{Gauss ini}
\end{align}
Here we have used $\tilA(t=0)=\sqrt{B_0}$ and $\delt\tilA(t=0)=0$. 
To search for physical unstable modes, we must solve the EOM from a initial condition which is consistent with the Gauss's law, Eq.~\eqref{Gauss ini}.
In this case, the physical degrees of freedom are $9+9-3=15$, and the physical fundamental matrix must be constructed by 15 independent physical solutions.		
The time evolution of an arbitrary physical mode is given by a linear combination of these solutions.  
So obtained growth rates do not depend on the choice of the initial fundamental matrix
as long as it consists of physical independent solutions.
				
In the following we give an example with $p_x=p_y=0$.
The generalization to $p_x, p_y\neq 0$ is straightforward.
Then, Eq.~\eqref{Gauss ini} becomes
\begin{align}
ie^1_zp_z/\sqrt{B_0}  &= -e^3_x ,\label{Gauss pz} \\
ie^2_zp_z/\sqrt{B_0}  &= e^3_y ,\\
ie^3_zp_z/\sqrt{B_0}  &= e^1_x-e^2_y .
\end{align}
For simplicity, we concentrate on $a^1_z$ and $a^3_x$ which satisfy the following equations of motion,
\begin{align}
\frac{d^2}{dt^2}\begin{pmatrix} a^1_z \\ a^3_x \end{pmatrix}
		  &= - \begin{pmatrix} \tilA^2   & i\tilA p_z             \\
								    -i\tilA p_z  & p_z^2 + \tilA^2  	\end{pmatrix}
				\begin{pmatrix} a^1_z \\ a^3_x \end{pmatrix}.
\label{EOM 1z-3x}
\end{align}
For $p_z\neq 0$, one initial condition which satisfies Eq.~\eqref{Gauss pz} is given by
\begin{align}
\Psi(t=0)
&=
\begin{pmatrix} 
1 & 0 & 0 & 0      \\
0 & 1 & 0 & 0      \\
0 & 0 & 1 & i\sqrt{B_0}/p_z      \\
0 & 0 & -ip_z/\sqrt{B_0} & 1
\end{pmatrix}\notag \\
&\equiv
\begin{pmatrix} 
\Phi(t=0) & * \\
* & 1
\end{pmatrix}\ ,
\end{align}
whose basis set is $\{a^1_z, a^3_x,e^1_z, e^3_x\}$.
The $3\times3$ matrix $\Phi$ consists of three independent modes $a^1_z, a^3_x$ and $e^1_z$.
After solving Eqs.~\eqref{EOM 1z-3x}, we get $\Psi(T)$ and hence $\Phi(T)$.
This is the way to construct the monodromy matrix for constraint systems.

In practice,
the fluctuations in the unphysical channels are found to show no unstable behavior and we can obtain the growth rate of the physical modes by using the initial condition $\Phi(0) = 1$.
We will discuss this point later.

\section{Instability bands of Yang-Mills equation}\label{sec:band}
\subsection{Global band structure}
\begin{figure}[bth]
\PSfig{8.5cm}{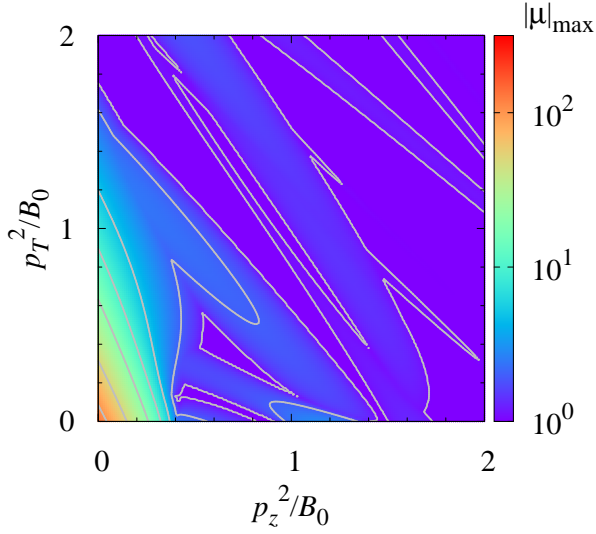}
\caption{
The contour map of the instability bands of classical Yang-Mills equation under oscillating color magnetic fields.
The contour lines (white lines)
stand for $|\mu|_{\rm max}=100,50,20,10,5,2,1.02$ from top to bottom.
The maximum characteristic multiplier is calculated as a function of both $p_z, p_T$, where $p_T = \sqrt{p_x^2 + p_y^2}$ is a transverse momentum.
Both $p_z$ and $p_T$ are rescaled by the initial strength of the background color magnetic field $B_0$.
The dominant instability band in the low momentum region has an anisotropy.
Transverse momentum direction of the dominant band is broader than that of $p_z$-direction which extends up to $p_T^2/B_0 = 1.75$ while the $p_z$ direction up to $p_z^2/B_0 = 0.81$.
}
\label{map}
\end{figure}

We determine the characteristic multipliers of the classical Yang-Mills equation by  numerical calculations.
Due to the axial symmetry along $z$-direction, the characteristic multipliers are the function of $p_z$ and $p_T = \sqrt{p_x^2 + p_y^2}$.
We calculate the maximum characteristic multiplier as a function of momentum $(p_z, p_T)$ by solving Eqs.~\eqref{Hill4} and~\eqref{Hill5}.
In Fig.~\ref{map}, we show the contour map of the instability bands, where both $p_z$ and $p_T$ are rescaled by the initial strength of the background color magnetic field $B_0$.
This contour map is obtained solely on the basis of Floquet theory, and thus the underlying nature of these instabilities is the parametric instability.

From Fig.~\ref{map}, we find that the instability band in the low momentum region around $p^2/B_0 \sim 0$ has an anisotropic shape and the large growth rate.
At $(p_z,p_T)=(0,0)$, the characteristic multiplier takes the maximum value, $|\mu|_\mathrm{max}=129$.
There are also other bands whose peak multipliers are less than about 5.
The broad instability region in the $p_T$-direction extends up to $p_T^2/B_0 \simeq 1.75$, while the instability region in the $p_z$-direction extends up to $p_z^2/B_0 = 0.81$.
The unstable region with $|\mu|_\mathrm{max}>5$ extends in the longitudinal and transverse momentum range, $p_z^2/B_0 < 0.37$ and $p_T^2/B_0 < 1.19$, respectively.
It means that the fluctuations is amplified by a factor 5 after a period of background field in the region.
If the strength of the background color magnetic field is scaled by the saturation momentum, the instability boundaries discussed here lies approximately at $p^2 \sim B \sim Q_{\rm s}^2$.
It is worth emphasizing that we find instabilities in both low momentum region $p \ll Q_{\rm s}$ and high momentum region $p \sim Q_{\rm s}$.
It should be noted that $|\mu|_\mathrm{max}\geq1$ is always satisfied as noted in Sec.~\ref{sec:Floquet}.

\subsection{Band structure for $p_T=0$ and $p_z=0$}
Now, let us try to understand the origin of the instabilities present in some other bands semi-analytically. 
For this purpose,
we consider two particular limits $p_T=0$ and $p_z=0$ of linearized EOMs of fluctuations Eqs.~\eqref{Hill4} and~\eqref{Hill5}, then we get simpler equations.
For $p_T=0$, Eqs.~\eqref{Hill4} and~\eqref{Hill5} are decomposed into two equations whose coefficients are $\Omega^2_B, \Omega^{*2}_C$ and $\Omega^2_A, \Omega^2_C$, respectively.
Similarly, for $p_z=0$, we can decompose Eqs.~\eqref{Hill4} and~\eqref{Hill5} into two independent equations whose coefficients are $\Omega^2_E, \Omega^2_G$ and $\Omega^2_D, \Omega^2_F$, respectively.
See Appendix~\ref{sec:Hill} for the explicit forms of these matrices and their relations.
These decomposed equations are easier to treat since their ranks are at most three so that they give insight into the whole band structure.

\begin{figure}[bth]
\PSfig{8.5cm}{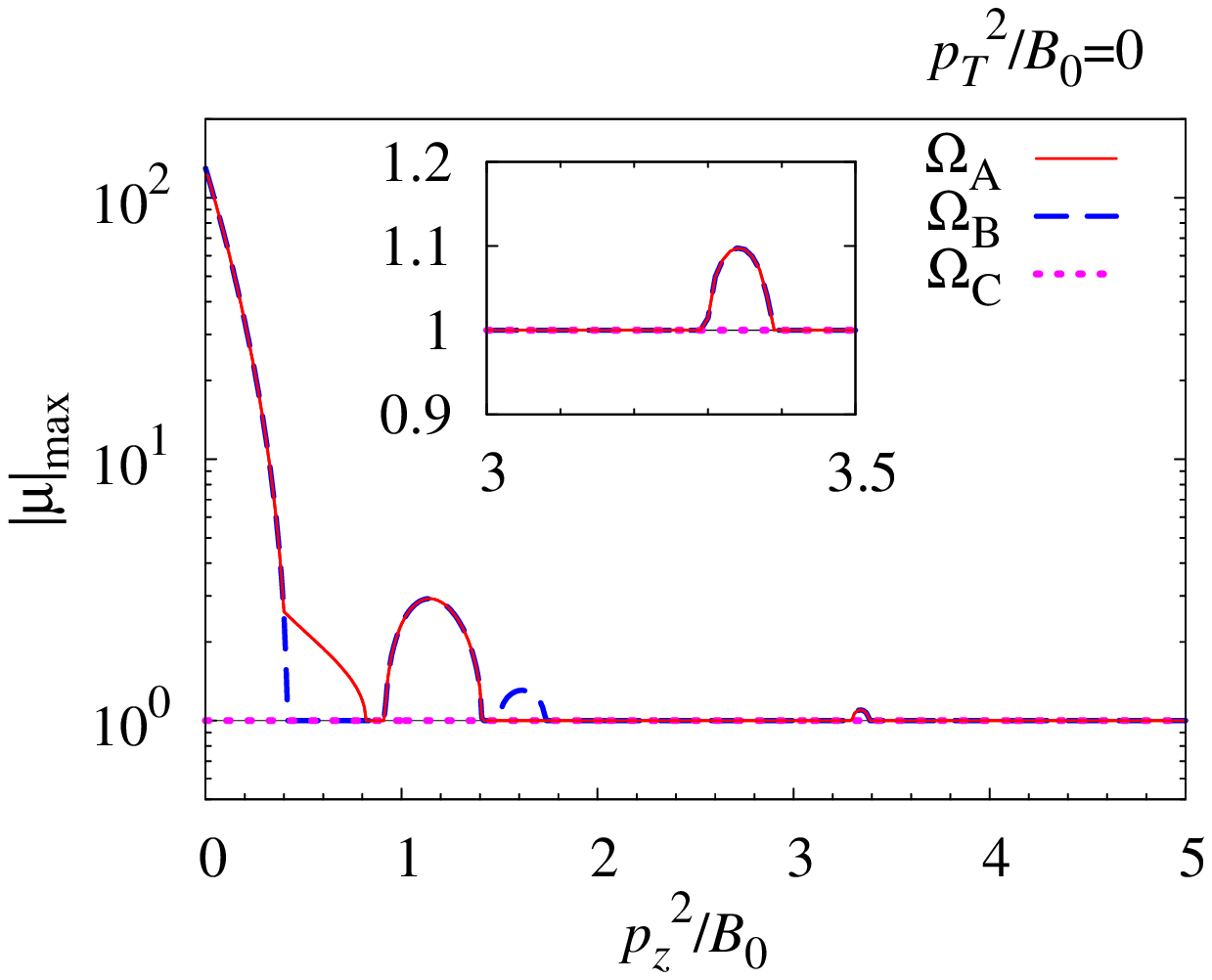}\\
\PSfig{8.5cm}{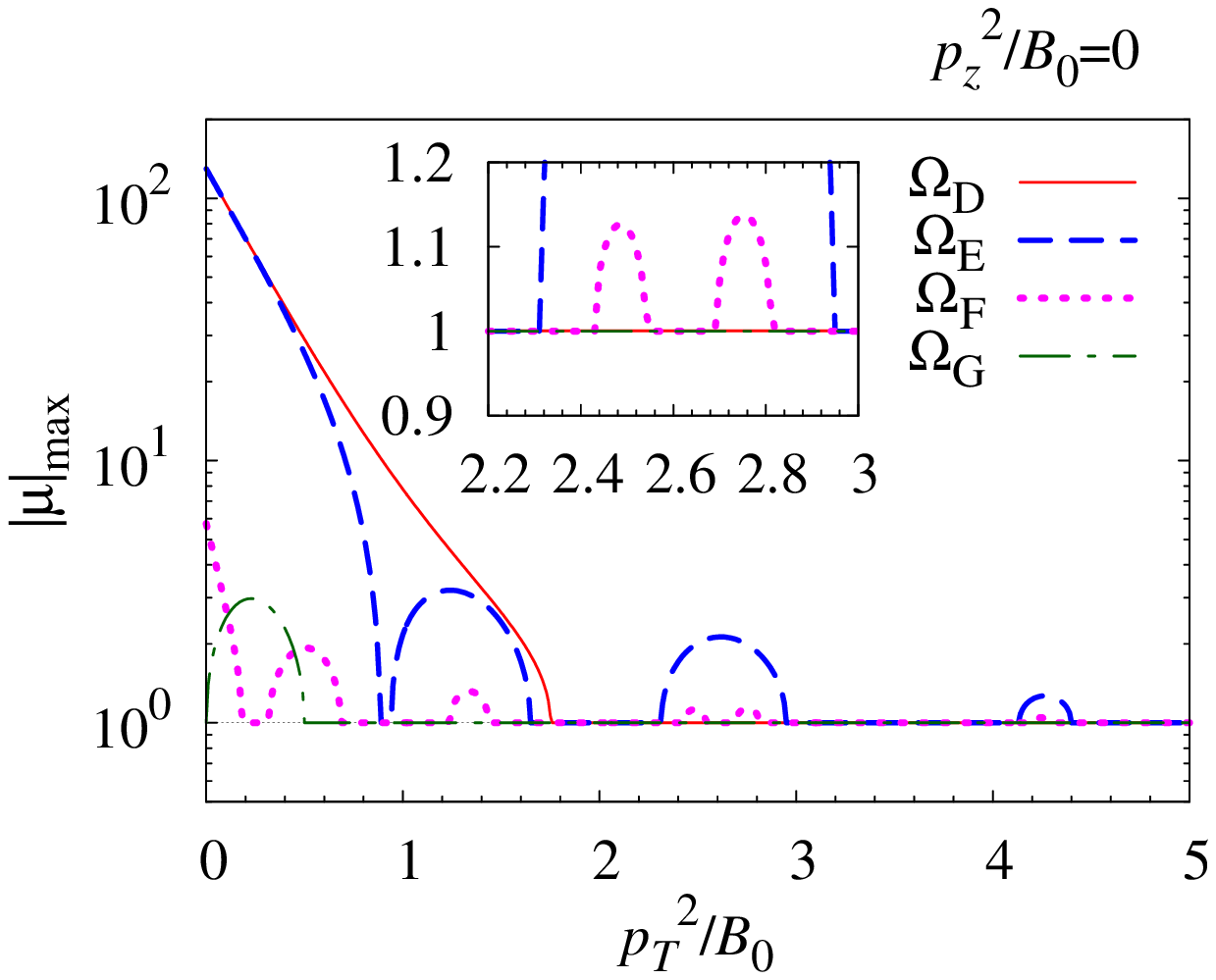}
\caption{
(top) Instability bands for $p_T=0$.
The lowest instability band of $\Omega^2_A$ system (red solid line) and $\Omega^2_B$ system (blue dashed line) located at $0 \leq p_z^2/B_0 \leq 0.41$ have the largest growth rate.
This dominant instability band is consistent with that found in~\cite{Berges12a}.
(bottom) Instability bands for $p_z=0$.
The lowest instability band of $\Omega^2_D$ system (red solid line) and $\Omega^2_E$ system (blue dashed line) have the largest growth rate.
They have broader band and reach $p_T^2/B_0 = 1.75$ and $p_T^2/B_0 = 0.88$, respectively.
The magenta dotted line and the green dot-dashed line stand for the bands of $\Omega^2_F$ and $\Omega^2_G$.
}
\label{band}
\end{figure}

Figure~\ref{band} shows the band structure in $p_T=0$ and $p_z=0$-regions.
Each line shows the maximum value of characteristic multipliers $|\mu|$ obtained from the decomposed equation with the matrix $\Omega^2_I (I=A, B, \ldots G)$.

For $p_T = 0$, the fluctuation fields $a^1_x, a^2_y, a^3_z, a^1_y$ and $a^2_x$ may be unstable and have larger growth rate than other components.
Almost all of these modes are related to $B^3_z$, and may modify the background color magnetic field.
The dominant instability bands of $\Omega^2_A$ and $\Omega^2_B$ systems range from $p_z^2/B_0 = 0$ to $p_z^2/B_0 = 0.81$ and to $p_z^2/B_0 = 0.41$, respectively.
Their growth rates become larger in the smaller momentum region.
This dominant instability band is consistent with that found in~\cite{Berges12a}.

The lower panel of Fig.~\ref{band} shows the instability bands for $p_z = 0$. 
In this case, $\Omega^2_D$ and $\Omega^2_E$ systems show the largest growth rate where the corresponding fluctuation fields are $a^a_x$ and $a^a_y$.
The ranges of the dominant instability band of these systems are broader than that of $\Omega^2_A$ and $\Omega^2_B$ systems.
The bands of $\Omega^2_D$ and $\Omega^2_E$ systems extends up to $p_T^2/B_0 = 1.75$ and $p_T^2/B_0 = 0.88$, respectively.
$\Omega^2_F$ and $\Omega^2_G$ systems also show instability, but with smaller growth rates and narrower ranges; the first and second instability bands of $\Omega^2_F$ system range from $p_T^2/B_0 = 0$ to $p_T^2/B_0 = 0.18$ and $p_T^2/B_0 = 0.48$ to $p_T^2/B_0 = 0.68$. The instability band of $\Omega^2_G$ system is identical to that of Lam\'e\rq{}s equation with $\epsilon = 1$, Eq.~\eqref{Lame1}.

\subsection{Effective reduction of EOMs}
In this subsection, we discuss the mathematical origin of the band structure.
Suppose that eigenvalues of a coefficient matrix $\Omega^2_I\,(I=A,B,\dots G)$ are given by $\omega^2_{I1},\dots,\omega^2_{Im},$ and 
$\Omega^2_I$ is diagonalized by a unitary matrix $U_I$;
\begin{align}
U_I^{-1}(t)\Omega^2_I(t)U_I(t) = \diag(\omega^2_{I1},\dots,\omega^2_{Im}) .
\end{align}
In general, $U_I$ depends on time because $\Omega^2_I$ includes the background gauge field $\tilA(t)$.
Multiplying an original EOM of fluctuations by $U_I^{-1}(t)$ from left, we get
\begin{align}
U_I^{-1}(t)\frac{d^2}{dt^2}\va 
=
-\diag(\omega^2_{I1},\dots,\omega^2_{Im}) U_I^{-1}(t) \va ,
\end{align}
where $\va$ is a corresponding gauge field.
For instance, $\va = (a^1_x, a^2_y, a^3_z)$ for $I=A$.
Therefore, if and only if $U_I$ is constant, the equation is decomposed into decoupled ones characterized by the eigenvalues of $\Omega^2_I$;
\begin{align}
\frac{d^2}{dt^2}\va^\prime
&=
-\diag(\omega^2_{I1},\dots,\omega^2_{Im}) \va^\prime , \\
\va^\prime &= U_I^{-1}\va .
\end{align}

\begin{figure*}[bth]
\PSfig{5.5cm}{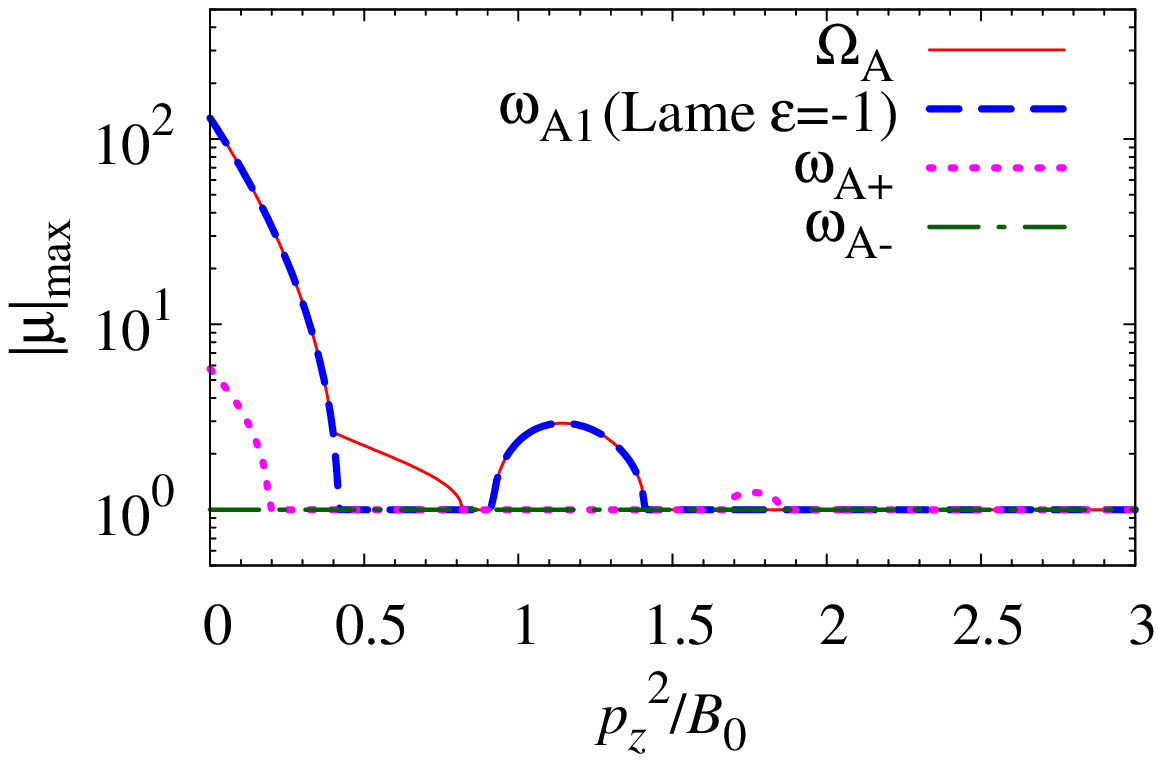}
\PSfig{5.5cm}{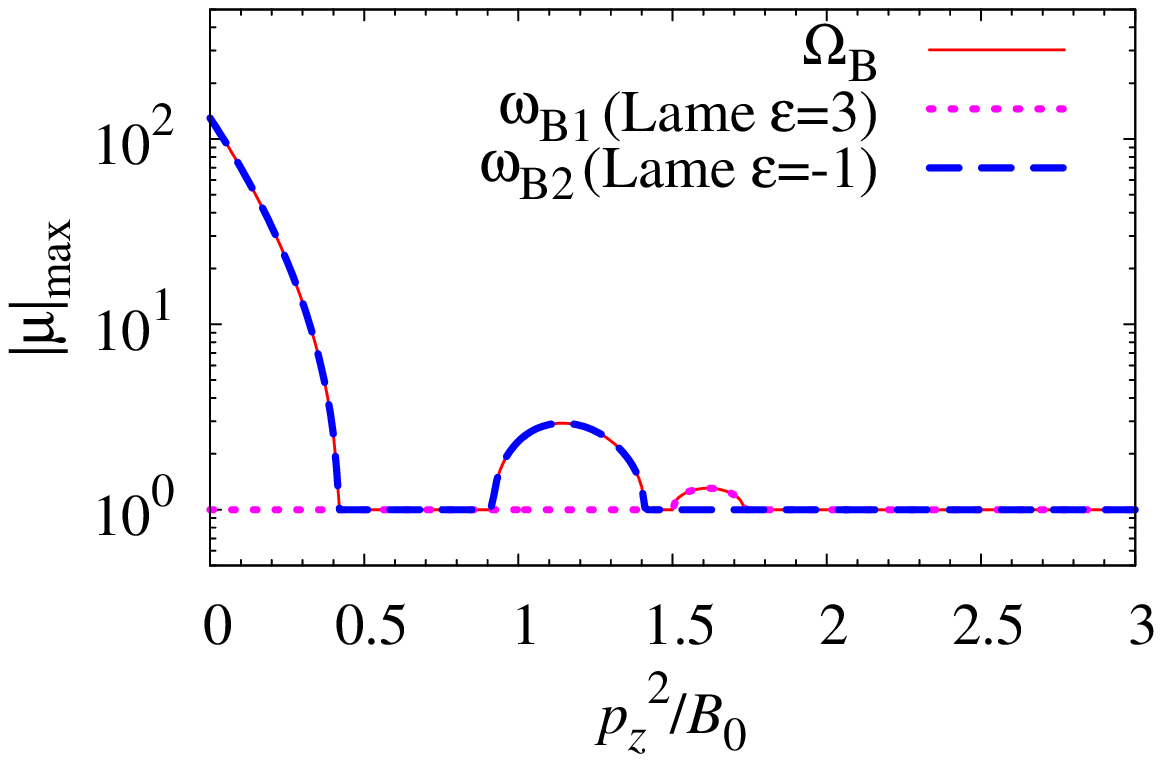}
\PSfig{5.5cm}{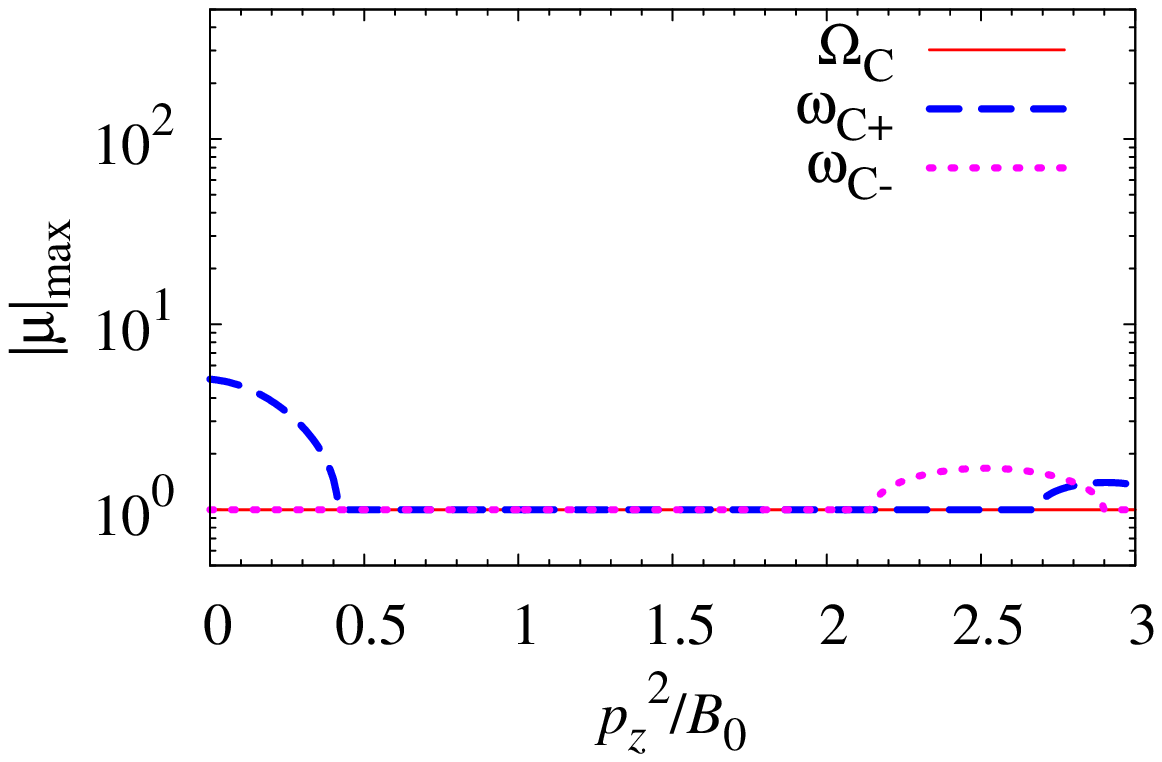}
\PSfig{5.5cm}{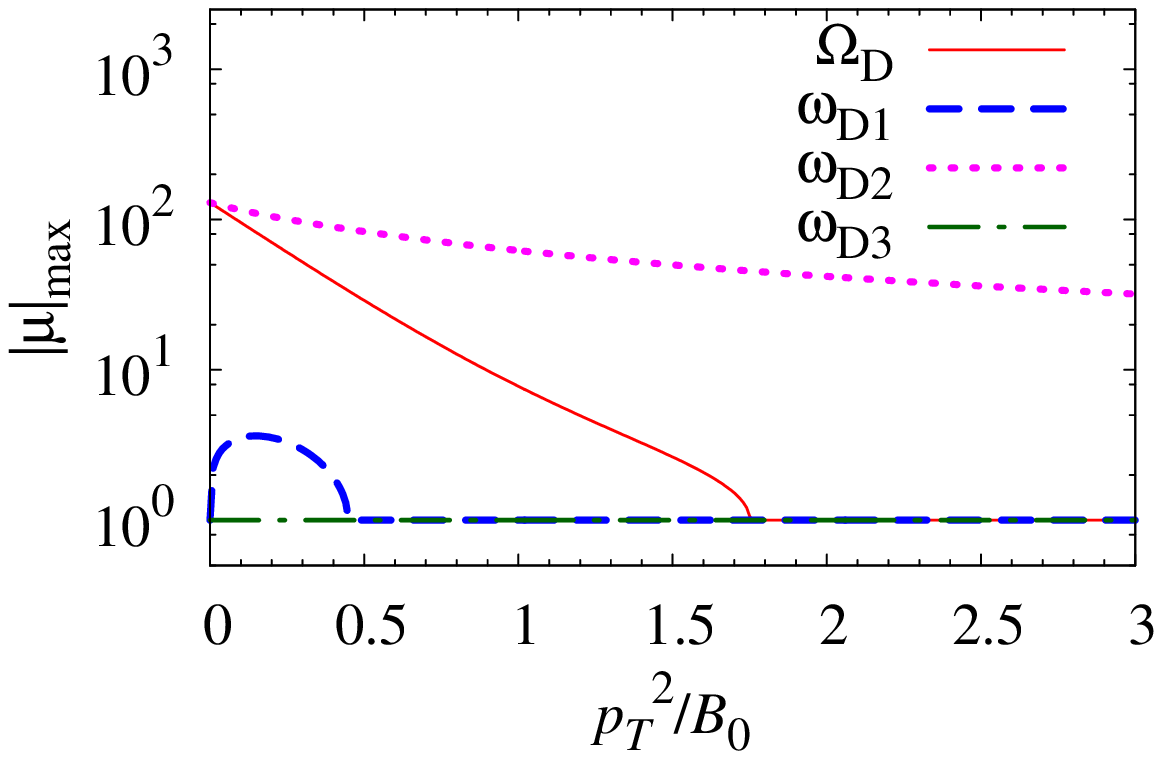}
\PSfig{5.5cm}{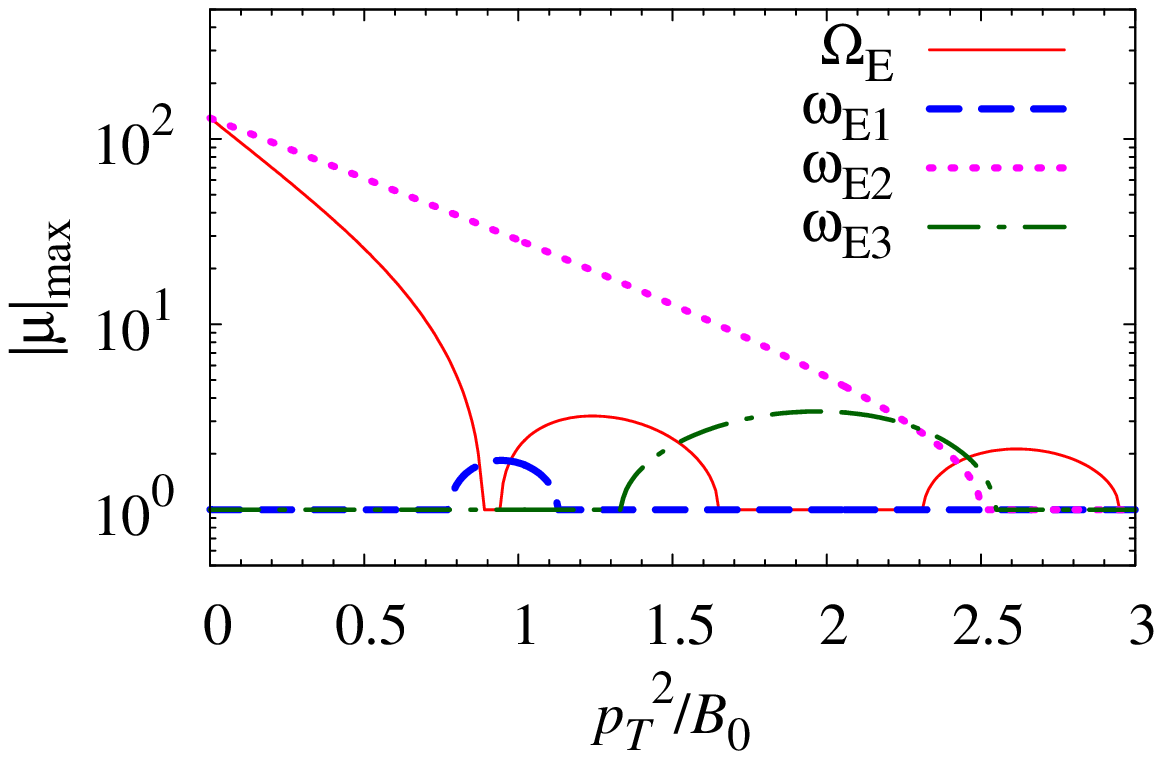}
\PSfig{5.5cm}{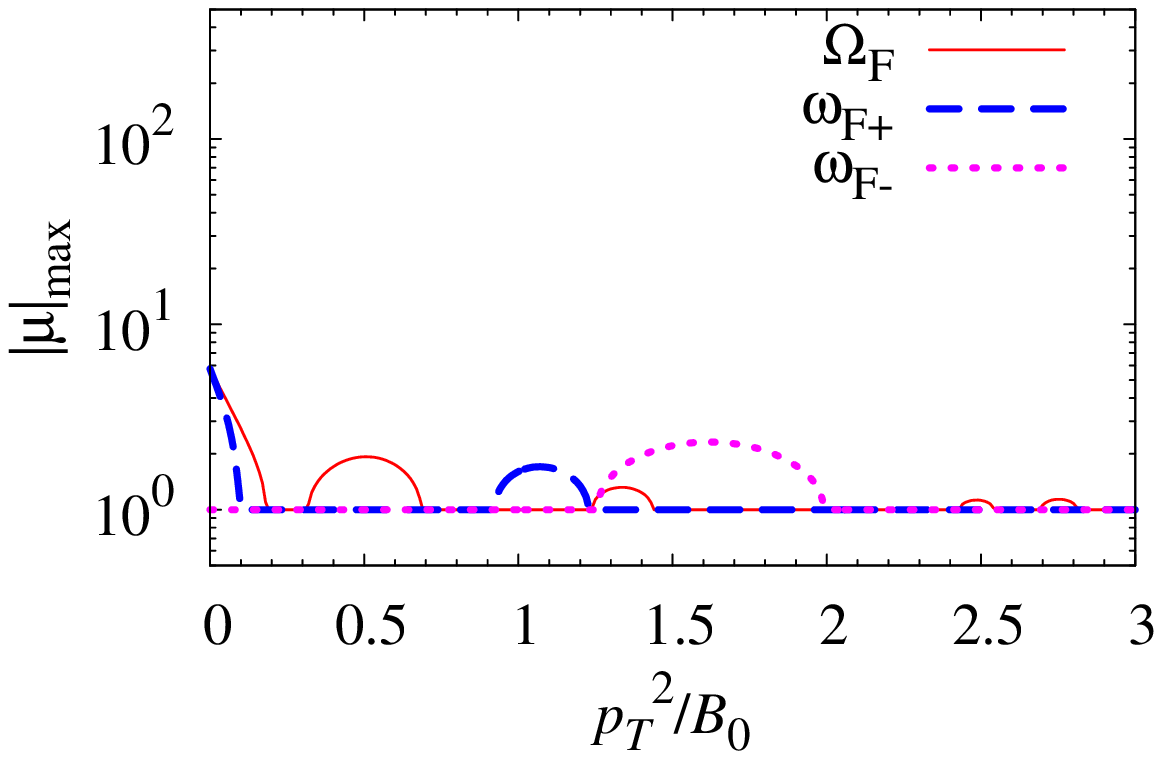}
\caption{
The comparison of instability bands of original EOMs (multi-component Hill\rq{}s equation with $\Omega_I^2$) and single component Hill\rq{}s equations (denoted by $\omega^2_{Im}$). 
(top) $\Omega_A^2$ system is effectively reduced to three single component Hill\rq{}s equations and one of them is Lam\'e\rq{}s equation with $\epsilon = -1$.
The instability bands of the Lam\'e\rq{}s equation are denoted by blue dashed line.
This result shows that most bands of the original EOM including the first band are well reproduced by Lam\'e\rq{}s equation.
$\Omega^2_B$ system is exactly reduced to two types of Lam\'e\rq{}s equations.
They also lead quite unstable behavior in low momentum region.
For $\Omega^2_C$ system, two Hill\rq{}s equations fail to reproduce the original bands.
(bottom) The change of instability bands is drastic in $\Omega^2_D, \Omega^2_E$ systems.The time dependence of $U_I$ makes instability regions considerably narrower for these equations.
}
\label{resAB}
\end{figure*}

Only the $\Omega^2_B$ system (EOM for $a^1_y$ and $a^2_x$) satisfies the above condition.
In this case, corresponding unitary matrix is given by
$U_B=1/\sqrt{2}\begin{pmatrix}1 & 1 \\ 1 & -1\end{pmatrix}$
and actually does not depend on time.
Eigenvalues of $\Omega^2_B$ are given by $\omega^2_{B1} = p_z^2 + 3\tilA^2$ and  $\omega^2_{B2} = p_z^2 - \tilA^2$.
As a result, we find two types of Lam\'e\rq{}s equations, Eqs.~\eqref{Lame3} and~\eqref{Lame-1}, and hence some parts of instability bands for $p_T=0$ are exactly described by Lam\'e\rq{}s equations.
Our numerical calculation confirms this point and the result is shown in the upper panel of Fig.~\ref{resAB}.
The first and second instability bands of $\Omega^2_B$ system (red solid line)
are found in the momentum range of $0 \leq p_z^2/B_0 = 0.41$ and $0.91 < p_z^2/B_0 < 1.42$, respectively.
They are identical to the instability bands of Lam\'e\rq{}s equation with $\epsilon = -1$, Eq.~\eqref{Lame-1} (blue dashed line). The third instability band ranges from $p_z^2/B_0 = 3/2$ to $p_z^2/B_0 = \sqrt{3}$ and this is identical to the band of Lam\'e\rq{}s equation with $\epsilon = 3$, Eq.~\eqref{Lame3} (magenta dotted line). The instability bands of Lam\'e\rq{}s equations are also depicted in Fig.~\ref{Lames}.

If $U_I(t)$ varies slowly in time, EOM of fluctuations are effectively reduced 
to a single component Hill\rq{}s equation; 
$\ddot a_{Im} = -\omega_{Im}^2 a_{Im}$.
For example, such an effective reduction seems to occur in the $\Omega^2_A$ system (EOM for $a^1_x, a^2_y$ and $a^3_z$). 
In fact, the eigenvalues of $\Omega_A^2$ are given by
\begin{align}
\omega_{A1}^2 &= p_z^2 - \tilA^2 ,
\label{1x2y3zdecouple1}\\
\omega_{A\pm}^2 
&= 
\half \lp p_z^2 + 2\tilA^2 \pm \sqrt{p_z^4 + 6p_z^2\tilA^2 + \tilA^4} \rp ,
\label{1x2y3zdecouple2}
\end{align}
and one of the resultant single component equation has the form of Lam\'e\rq{}s equation with $\epsilon = -1$.
We compare the instability bands of the original EOM and three Hill\rq{}s equations.
The upper panel of Fig.~\ref{resAB} shows that the instability bands of $\Omega^2_A$ system (red solid line) is well described by Lam\'e\rq{}s equation (blue dashed line).
The other two Hill\rq{}s equations show a weaker instability than Lam\'e\rq{}s equation.
While three Hill\rq{}s equations do not completely reproduce the band structure of $\Omega^2_A$ system, the bulk structure is reproduced and the time dependence of $U_A(t)$ is considered to be small.

The dominant instability band in $p_T=0$-region is seemingly reminiscent of the Nielsen-Olesen instability as discussed in~\cite{Berges12a}.
However, the dominant band is described by Lam\'e\rq{}s equation with $\epsilon = -1$, Eq.~\eqref{Lame-1}, in both $\Omega_A$ and $\Omega_B$ systems.
Therefore, we conclude that the Nielsen-Olesen instability-like behavior actually comes from parametric instability.

It should be noted that, in contrast to $\Omega^2_A$ and $\Omega^2_B$ systems, the other EOMs cannot be regarded as a set of single component Hill\rq{}s equations.
For the $\Omega^2_C$ system, the two Hill\rq{}s equations have some instability bands while the original EOM has no instability band.
For $p_z = 0$, there is a substantial difference in the band structure between the original EOM and a set of single component Hill\rq{}s equations.
Actually, the lower panel of Fig.~\ref{resAB} shows that Hill\rq{}s equations fail to reproduce qualitative behaviors of instability bands of the original EOMs.
In this sense, even for the dominant instability band, the interpretation of band structure in terms of Lam\'e\rq{}s equations is valid in very limited cases.

\subsection{Implication to unphysical sector}
\begin{figure}[bth]
\PSfig{8.5cm}{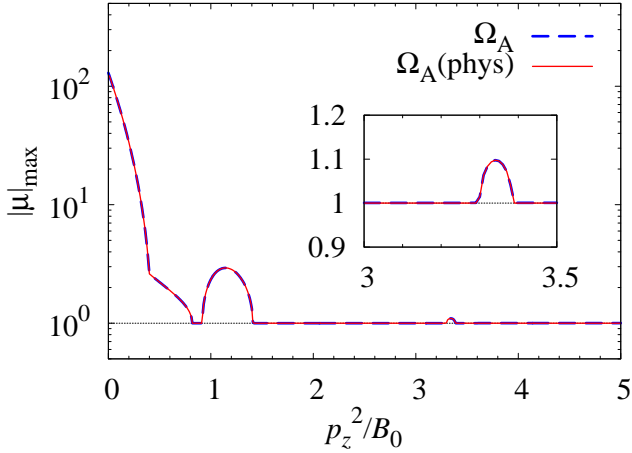}
\caption{
Comparison of instability bands obtained with a physical initial condition and a unit matrix initial condition.
The red solid line denotes the instability bands of $\Omega^2_A$ system calculated by physical initial condition.
The blue dashed line results from the unit matrix initial condition.
}
\label{gauss check}
\end{figure}

In our calculation,
we have constructed the physical initial condition which is consistent with the Gauss's law in order not to pick up the instabilities of unphysical modes.
We present the construction of physical initial condition introduced in Sec.~\ref{app to YM}. 
We have also confirmed that the solution at $t=T$ satisfy Gauss's law numerically.
In practice, we can start our calculation from the simple initial conditions which do not necessarily fulfill the Gauss's law.
We compare two calculations using physical initial condition and unit matrix initial condition given by $\Phi(0) = 1$.
Figure~\ref{gauss check} shows that the maximum characteristic multipliers of $\Omega^2_A$ system using two different initial conditions.
In principle, the calculation resulted from the unit matrix initial condition can contain the contributions from unphysical modes, however, the two results completely agree with each other.
We have also confirmed that all unphysical modes only appears as stable modes.
These results mean that the unphysical sector does not contain unstable modes, so all instabilities we get are physical ones.
This property simplifies the Floquet analysis in a practical sense because we can set the initial fundamental matrix as a unit matrix which is simpler than the physical fundamental matrix. 
We also find this property for full band structure.
\section{Conclusions}
\label{sec:summary}
We have studied the nature of instabilities of the classical Yang-Mills equation under the time-dependent homogeneous color magnetic field in the linear regime.
The background color magnetic field considered in this article is realized by the Berges-Scheffler-Schlichting-Sexty (BSSS)~\cite{Berges12a} gauge configuration which does not have spatial dependence, and thus a transverse momentum is well defined.
This gauge configuration gives color electric fields in the $x$ and $y$-directions as well.
It is known that there exists a dominant instability region at $p_z^2/B_0 < 1$ under the BSSS gauge field.
It is also suggested that there is a sub-dominant instability band at $p_z^2/B_0 > 1$.
The $p_z$-dependence of the growth rate in the dominant instability region behaves as if it is caused by the Nielsen-Olesen instability.
However, the growth rate must also depend on $p_T$, and the band structure in $(p_z,p_T)$-plane was not investigated so far.
We have made linear analysis in order to search for the complete band structure and to reveal the nature of the instabilities.
The Floquet analysis is best suited to determine instability boundaries and we have applied it to the Yang-Mills theory.

We have found that there is a broad instability band which have the maximum growth rate around zero momentum region in the $(p_z,p_T)$-plane.

For $p_T=0$, the band reaches $p_z^2/B_0 \simeq 0.41$ and this result is consistent with Ref.~\cite{Berges12a}.
We have also found many other sub-dominant instability bands.
Moreover, we have rediscovered the Nielsen-Olesen instability like behavior of the dominant instability band in the small $p_T$ region.
For $p_z=0$, the instability band which has the maximal growth rate locates around zero momentum region.
The dominant instability band is broader than that of $p_T=0$ case, and it extends up to $p_T^2/B_0 \simeq 1.75$.
However, in the system considered in this article, the background gauge field does not form Landau levels, and thus the present instability has nothing to do with the instability of the lowest Landau level, namely the Nielsen-Olesen instability.

We have investigated the mathematical origin of the seemingly Nielsen-Olesen like behavior.
As a result, the dominant instability band which is reminiscent of Nielsen-Olesen instability is dominated by Lame\rq{}s equation whose dispersion relation is effectively given by $\omega = \sqrt{p_z^2 - \tilA^2(t)}$, due to the effective reduction of the original EOMs.
The present dominant instability band in the small $p_T$ region shares the common feature as the Nielsen-Olesen instability in the sense that the negative eigenvalues of $\Omega^2_A$ and $\Omega^2_B$ appear from the combination of the polarization modes, $a^1_x-a^2_y$ and $a^1_y-a^2_x$.
This feature comes from the anomalous Zeeman effect for the spin 1 systems.
At the same time, the present instability is not caused by the instability of the lowest Landau level, but by bunch of unstable modes with a band structure extending to continuous transverse momenta,
and it is not literally the Nielsen-Olesen instability. 
Therefore, we have concluded that the origin of all these instabilities considered here is regarded as the result of parametric instability.

It is instructive to consider whether the parametric instability considered in this article persist or not when the nonlinear interaction between gluons becomes important.
When the background field loses its energy and the amplitude of gauge fluctuation becomes comparable to the background field $\sqrt{B_0}$, the nonlinear effects cannot be negligible,  and thus the instability bands defined in the linear regime lose their validity.
It is expected that the signal of the resonance bands in the high momentum region may be quite weak, since their growth rates are much smaller than the growth rate around the zero momentum.
However, for $p_z = 0$, the growth rate varies gradually as a function of $p_T$ and the range of the instability band is broad.
Even for $p_T^2/B_0 \sim 1$, the growth rate is not so small.
Such a momentum dependence of the growth rate is unique to the parametric instability,  and the behavior of the growth rate is expected to be seen in full numerical calculations.
These studies are in progress.

From a phenomenological point of view, it is also important to consider how the parametric instability is affected by longitudinal expansion and inhomogeneity of the background field.
In a longitudinal expanding geometry, strength of the background color magnetic field 
would damp and the growth rates of the unstable modes become small.
Thus, longitudinal expansion would make the signal of the parametric instability weak.
Spatial inhomogeneity of the background field is also expected to suppress the parametric instability.
If the scale of inhomogeneity is given by the saturation scale $Q_{\rm s}$, the relevant modes of the dynamics have $p^2 \gtrsim Q_{\rm s}^2 \simeq B_0$.
Indeed, the parametric instability in the region is weaker than that in lower momentum region as $p^2 \lesssim Q_{\rm s}^2 \simeq B_0$. 
However, as we have mentioned above, the growth rates of unstable modes around $p^2 \sim Q_{\rm s}^2 \simeq B_0$ are not so small and they may affect the early stage dynamics. 
The qualitative discussions on these points are kept for a future work.

\section*{Acknowledgment}
We would like to thank J. Berges, K. Itakura, B. M\"{u}ller, S. Schlichting and R. Venugopalan for useful discussions. 
The authors also would like to thank the participants of the Nishinomiya Yukawa Memorial \& YIPQS workshop on ``New Frontiers in QCD 2013'' (YITP-T-13-05) for useful discussions.
This work was supported in part by 
the Grants-in-Aid for Scientific Research from JSPS
 (Nos. 20540265, 
 23340067, 
 24340054, 
 24540271
),
the Grants-in-Aid for Scientific Research on Innovative Areas from MEXT
 (No. 2004: 23105713, and No. 2404: 24105001, 24105008),
 by the Yukawa International Program for Quark-Hadron Sciences,
 by a Grant-in-Aid for the global COE program
``The Next Generation of Physics, Spun from Universality and Emergence'' from MEXT.
S.T. is supported by the Grant-in-Aid for JSPS fellows (No.26-3462) and JSPS Strategic Young Researcher Overseas Visits Program for Accelerating Brain Circulation (No.R2411).
T.K. is supported by the Core Stage Back Up program in Kyoto University.

\appendix

\section{CYM equation on $p_T=0$ and $p_z=0$}
\label{sec:Hill}
\subsection{Notations}
In this appendix, we give the specific form of the linearized equations for the fluctuation of gauge fields.
The coefficient matrix of EOM for fluctuations $a^a_i$ is denoted by $\Omega^2$ as in Eq. \eqref{Hill}.
Without loss of generality, we can take $p_y=0$ due to the rotational symmetry of  transverse direction and the EOM is decomposed into two independent sectors,
\begin{align}
\ddot{a}_\alpha &= - [\Omega^2_4]_{\alpha\beta} a_\beta, \quad 
a_\alpha = (a^1_y,a^2_x,a^2_z,a^3_y), \\
\ddot{a}_A &= - [\Omega^2_5]_{AB} a_B, \quad
a_A = (a_x^1, a_z^1, a_y^2 , a_x^3, a_z^3),
\end{align}
where each coefficient matrix is given by
\begin{align}
\Omega^2_4 
&= 
\begin{pmatrix}
p_x^2 + p_z^2 + \tilA^2 & 2\tilA^2 & 0 & -2ip_x\tilA \\
2\tilA^2 & p_z^2 + \tilA^2 & -p_xp_z & -ip_x\tilA \\
0 & -p_xp_z & p_x^2 + \tilA^2 & -ip_z\tilA \\
2ip_x\tilA & ip_x\tilA & ip_z\tilA & p_x^2 + p_z^2 + \tilA^2
\end{pmatrix},
\label{system4}\\
\Omega^2_5
&= 
\begin{pmatrix}
p_z^2 & -p_xp_z & -\tilA^2 & 0 & ip_z\tilA \\
-p_xp_z & p_x^2 + \tilA^2 & 0 & ip_z\tilA & -2ip_x\tilA \\
-\tilA^2 & 0 & p_x^2 + p_z^2 & -ip_x\tilA & -ip_z\tilA \\
0 & -ip_z\tilA & ip_x\tilA & p_z^2 + \tilA^2 & -p_xp_z \\
-ip_z\tilA & 2ip_x\tilA & ip_z\tilA & -p_xp_z & p_x^2 + 2\tilA^2
\end{pmatrix}.
\label{system5}
\end{align}
When we consider certain limits, the EOM becomes simplified.
For the case $p_T=0$, the EOM is reduced to four equations,
\begin{align}
\frac{d^2}{dt^2}\begin{pmatrix} a_x^1 \\ a_y^2 \\ a_z^3 \end{pmatrix}
&= 
- \Omega^2_A \begin{pmatrix} a_x^1 \\ a_y^2 \\ a_z^3 \end{pmatrix}, \,
\frac{d^2}{dt^2}\begin{pmatrix} a^1_y \\ a^2_x \end{pmatrix} 
= 
- \Omega^2_B \begin{pmatrix} a^1_y \\ a^2_x \end{pmatrix}, \notag \\
\frac{d^2}{dt^2}\begin{pmatrix} a^1_z \\ a^3_x \end{pmatrix}
&= 
- \Omega^2_C \begin{pmatrix} a^1_z \\ a^3_x \end{pmatrix}, \,
\frac{d^2}{dt^2}\begin{pmatrix} a^2_z \\ a^3_y \end{pmatrix} 
= 
- \Omega^{*2}_C \begin{pmatrix} a^2_z \\ a^3_y \end{pmatrix},
\end{align}
where each coefficient matrices $\Omega^2_I$ are given by
\begin{align}
&\Omega^2_A
= 
\begin{pmatrix} p_z^2 & -\tilA^2 & i\tilA p_z   \\
					-\tilA^2 & p_z^2          & -i\tilA p_z \\
					-i\tilA p_z  & i\tilA p_z   & 2\tilA^2	\end{pmatrix} ,
\label{systemA} \\
&\Omega^2_B
= 
\begin{pmatrix} p_z^2 + \tilA^2 & 2\tilA^2          \\
				 	2\tilA^2          & p_z^2 + \tilA^2  	\end{pmatrix} ,
\label{systemB} \\
&\Omega^2_C
= 
\begin{pmatrix} \tilA^2   & i\tilA p_z             \\
					-i\tilA p_z  & p_z^2 + \tilA^2  	\end{pmatrix} .
\label{systemC}
\end{align}
Their eigenvalues are given by
\begin{align}
\omega_{A1}^2 &= p_z^2 - \tilA^2 ,\\
\omega_{A\pm}^2 &= 
\frac{1}{2}   \lp p_z^2 + 3\tilA^2 \pm \sqrt{ p_z^4 + 6p_z ^2\tilA^2 + \tilA^4} \rp ,\\
\omega_{B1}^2 &= p_z^2 + 3\tilA^2 ,\\
\omega_{B2}^2 &= p_z^2 - \tilA^2 ,\\
\omega_{C\pm}^2 &= 
\frac{1}{2}   \lp p_z^2 + 2\tilA^2 \pm \sqrt{ p_z^4 + 4p_z ^2\tilA^2} \rp,
\end{align}
where $\omega^2_{I1},\,\dots\,\omega^2_{Im}$ stand for the eigenvalues of  $\Omega^2_I$.
We use the explicit form of them in order to calculate the instability bands of single component Hill\rq{}s equations.
Note that $\omega_{A1}^2,\,\omega_{B1}^2$ and $\omega_{B2}^2$ correspond to Lam\'e\rq{}s equations which are the special case of Hill\rq{}s equation.

For the case $p_y=p_z=0$,
\begin{align}
\frac{d^2}{dt^2}\begin{pmatrix} a_x^1 \\ a_y^2 \\ a_x^3 \end{pmatrix}
&= - \Omega^2_D \begin{pmatrix} a_x^1 \\ a_y^2 \\ a_x^3 \end{pmatrix}, \,
\frac{d^2}{dt^2}\begin{pmatrix} a_y^1 \\ a_x^2 \\ a_y^3 \end{pmatrix}
= - \Omega^2_E \begin{pmatrix} a_y^1 \\ a_x^2 \\ a_y^3 \end{pmatrix}, \notag \\
\frac{d^2}{dt^2}\begin{pmatrix} a^1_z \\ a^3_z \end{pmatrix}
&= - \Omega^2_F \begin{pmatrix} a^1_z \\ a^3_z \end{pmatrix} , \,\,\,
\frac{d^2}{dt^2}a^2_z 
= - \Omega^2_G a^2_z ,
\end{align}
where each coefficient matrices $\Omega^2_I$ are given by
\begin{align}
\Omega^2_D 
&= 
\begin{pmatrix} 0          & -\tilA^2 & 0   \\
									-\tilA^2 & p_x^2          & -i\tilA p_x \\
									0  & i\tilA p_x   & \tilA^2	\end{pmatrix},
\label{systemD} \\
\Omega^2_E 
&= 
\begin{pmatrix} p_x^2 + \tilA^2     & 2\tilA^2 & -2i\tilA p_x    \\
									2\tilA^2 & \tilA^2     & -i\tilA p_x \\
									2i\tilA p_x  & i\tilA p_x   &p_x^2 + \tilA^2	\end{pmatrix},
\label{systemE} \\
\Omega^2_F  
&= 
\begin{pmatrix} p_x^2 + \tilA^2 & -2i\tilA p_x \\ 
								2i\tilA p_x   & p_x^2+2\tilA^2 \end{pmatrix},
\label{systemF} \\
\Omega^2_G 
		  &= (p_x^2 + \tilA^2) .
\label{systemG}
\end{align}
Their eigenvalues are given by
\begin{align}
&\omega_{Dm}^6 - (\tilA^2 + p_T^2)\omega_{Dm}^4
- \tilA^4\omega_{Dm}^2 
+ \tilA^6  = 0 ,\label{omegaD}\\
&\omega_{Em}^6 - (3\tilA^2 + 2p_T^2)\omega_{Em}^4 \\
&\quad\quad - (\tilA^4 + \tilA^2p_T^2 - p_T^4)\omega_{Em}^2 + 3\tilA^6 - \tilA^4p_T^2 = 0, \notag \label{omegaE}\\
&\omega_{F\pm}^2
=
\half \lp 3\tilA^2 + 2p_T^2 \pm \tilA \sqrt{\tilA^2 + 16 p_T^2} \rp .
\end{align}
$\omega^2_{Dm},\,\omega^2_{Em}$ are given as the solutions of Eqs.~\eqref{omegaD} and ~\eqref{omegaE}, respectively.
Of course, we can get the explicit form of them, but they are quite complicated.
Note that EOM for $a^2_z$ decouples from other components to be Lam\'e\rq{}s equation.

In summary, the coefficient matrices $\Omega_I^2$ hold following relations,
\begin{align}
\Omega^2 &= \diag{(\Omega_4^2, \Omega_5^2)}  \quad(\text{for all }p_z,p_T) ,\\
\Omega_4^2 &= 
\begin{cases}
\diag{(\Omega_B^2, \Omega_C^{*2})} & (p_T=0) \\
\diag{(\Omega_E^2, \Omega_G^2)} & (p_z=0)
\end{cases},\\
\Omega_5^2 &= 
\begin{cases}
\diag{(\Omega_A^2, \Omega_C^2)} & (p_T=0) \\
\diag{(\Omega_D^2, \Omega_F^2)} & (p_z=0)
\end{cases}.
\end{align}

\section{Instability boundaries of Lam\'e\rq{}s equation}\label{App:Lame}
\subsection{Multiple-scale analysis}\label{sec:multi}
The instability boundaries of Lam\'e\rq{}s equation are calculable in perturbative way  even when the closed form solutions are not available. 
The method discussed here is called multiple-scale analysis which is a kind of singular perturbation~\cite{Bender78}.
We consider following equation with dimensionless parameter $\lambda$, $\epsilon$.
\begin{align}
\ddot{f} + \lp \lambda + \epsilon \cn^2(t;k) \rp f = 0 .
\label{pLame}
\end{align}
If $\epsilon \ll 1$, we can treat \lq\lq{}external force term\rq\rq{} as a perturbation.
When $k=0$, Lam\'e\rq{}s equation comes down to Mathieu\rq{}s equation.
Lam\'e\rq{}s equation with $k=1/\sqrt{2}$ appears in scalar $\phi^4$-theory and Yang-Mills theory. 
In this section, we perform the multiple-scale analysis for arbitrary $k$.
Hereafter, we use abbreviations as, $K=K(k), \Kp=K(\sqrt{1-k^2})$ and $E=E(k)$.
$K(k)$ and $E(k)$ are complete elliptic integrals of the first kind and second kind, respectively.
Their explicit form are given by
\begin{align}
	K(k) &= \int^{\pi/2}_0 d\phi \frac{1}{\sqrt{1-k^2\sin^2\phi}}, \\
	E(k) &= \int^{\pi/2}_0 d\phi \sqrt{1-k^2\sin^2\phi} .
\end{align} 

By substituting the Fourier expansion for the elliptic function,
Eq.~\eqref{pLame} becomes
\begin{align}
\ddot{f}
+ \lp c + \frac{\epsilon}{k^2} \sum_{m=1}^\infty b_m \cos \frac{m\pi}{K}t \rp f = 0 ,\label{fourierlame}\\
c = \lambda + \frac{\epsilon b_0}{k^2} .
\end{align}
Here we use following formula,
\begin{align}
k^2\cn^2u 
	= \underbrace{k^2 - 1 + \frac{E}{K}}_{b_0} 
	+\sum_{m=1}^\infty 
	\underbrace{\frac{\pi^2}{K^2}\frac{m}{\sinh  m\pi\Kp/K}}_{b_m} \cos\frac{m\pi}{K}u .
\end{align}
We can perform multiple-scale analysis parallel to Mathieu's equation using Fourier expansion of elliptic function.
First, we simply expand $f$ as $f = f_0 + \epsilon f_1 + \dots$ in order to search for the starting points of instability boundaries.
The equation of order $\epsilon$ becomes
\begin{align}
&\ddot{f_1} + cf_1 \notag \\ 
&= 
- A_0\sum_{m=1}^\infty \frac{b_m}{2k^2}	 
e^{i\sqrt{c}} \lp e^{im\pi t /K} + e^{-im\pi t /K} \rp
+ \cc\, .
\end{align}
If resonance occurs, the solution will be strongly amplified and unstable.
The conditions for the resonance are given by
\begin{align}
c = \lp\frac{m\pi}{2K}\rp^2  \quad m^2=1,4,9,\dots .
\end{align}
This condition gives the starting point of the instability boundaries.
Therefore, it is convenient to expand $c$ as,
\begin{align}
c &= \lp\frac{m\pi}{2K}\rp^2 + \epsilon c_1 + \dots ,
\end{align}
and substitute it to Eq.~\eqref{fourierlame},
\begin{align}
\ddot{f}
+ \lb \lp\frac{m\pi}{2K}\rp^2  
+ \lp c_1 + \frac{1}{k^2} \sum_{m=1}^\infty b_m \cos \frac{m\pi}{K}t \rp \epsilon
\rb f = 0.
\label{pLame mlt}
\end{align}
Now we introduce a slow variable $\tau = \epsilon t$ and assume that $f$ is a function of $t$ and $\tau$.
This is the main idea of multiple-scale analysis.
We expand $f$ as,
\begin{align}
f &= f_0(t,\tau) + \epsilon f_1(t,\tau) + \cdots .
\end{align}
In contrast to the naive perturbation theory, the leading order solution has $\tau$ dependence.  
$f_0 = A(\tau)\exp\lp im\pi t/2K\rp + \cc$~. The equation of $\mathcal{O}(\epsilon)$ is also modified as follows,
\begin{align}
\frac{\partial^2 f_1}{\partial t^2} &+ \lp\frac{m\pi}{2K}\rp^2 f_1 \notag\\
=
&- \lp c_1A + \frac{b_m}{2k^2}	A^* + i\frac{m\pi}{K}\frac{dA}{d\tau} \rp  \exp \lp i\frac{m\pi}{2K}t \rp \nonumber\\
&- \sum_{m\neq n} \frac{b_n}{2k^2}	 A^* \exp \lp i\frac{(n-m/2)\pi}{K}t \rp  \nonumber\\
&- \sum_{n} \frac{b_n}{2k^2}	 A \exp \lp i\frac{(n+m/2)\pi}{K}t \rp
+\cc\,.
\label{NLO}
\end{align}
There is secular divergence due to the first term of the right hand side of Eq~\eqref{NLO}.
Thanks to the slow variable dependence of $A(\tau)$, we find the \lq\lq{}renormalization equation\rq\rq{} so as to remove the secular divergence, namely, $A(\tau)$ must satisfy the following relation,
\begin{align}
c_1A + \frac{b_m}{2k^2}A^* + i\frac{m\pi}{K}\frac{dA}{d\tau} = 0 . 
\end{align}
We can solve this equation by putting $A=B+iC$, then we get
\begin{align}
B(\tau) = \text{const.} \times \exp \lp \pm \frac{K}{m\pi} \sqrt{\frac{b_m^2}{4k^4} - c_1^2} \rp \tau .
\end{align}
Thus, we find that if $|c_1| <  b_m/2k^2$ the solution is unstable.
This gives the instability boundaries in the accuracy of $\mathcal{O}(\epsilon)$,
\begin{align}
c = 
\lp\frac{m\pi}{2K}\rp^2 
\pm \frac{\pi^2m}{2k^2K^2\sinh  m\pi\Kp/K }\epsilon 
+ \cdots .
\end{align}
Now it is clear that the instability boundaries are calculable up to arbitrary order of $\epsilon$ in a systematic way.
Here we note the explicit form of instability boundaries up to the order $\mathcal{O}(\epsilon^2)$ for the sake of completeness;
\begin{align}
&\lambda_\pm
= 
\lp \frac{m\pi}{2K} \rp^2 \\
&+
\lbb 
\frac{1}{k^2} - 1 - \frac{E}{k^2K} \pm \frac{\pi^2}{2k^2K^2}\frac{m}{\sinh m\pi\Kp/K} 
\rbb\epsilon \notag \\
&+ 
\frac{\pi^2 B_\mp(m)}{4k^4K^2}\epsilon^2 + \mathcal{O}(\epsilon^3), \notag \\
&B_\mp(m) 
=
\sum_{n=1, \, (n\neq m)}^\infty \frac{1}{\sinh n\pi\Kp/K} \\
&\times\lbb
\frac{1}{\sinh (n-m)\pi\Kp/K} \mp \frac{n}{n-m}\frac{1}{\sinh n\pi\Kp/K}
\rbb. \notag
\end{align}
For the case $k=1/\sqrt{2}$ and $m=1$, we get
\begin{align}
\lambda_+
&\simeq 
0.72 - 0.21\epsilon , \\
\lambda_-
&\simeq 
0.72 - 0.71\epsilon .
\end{align}
For the case $m=2$,
\begin{align}
\lambda_+
&\simeq 
2.87 - 0.44\epsilon , \\
\lambda_-
&\simeq 
2.87 - 0.48\epsilon - 0.04\epsilon^2  .
\end{align}

\begin{figure}[bthp]
\PSfig{8.5cm}{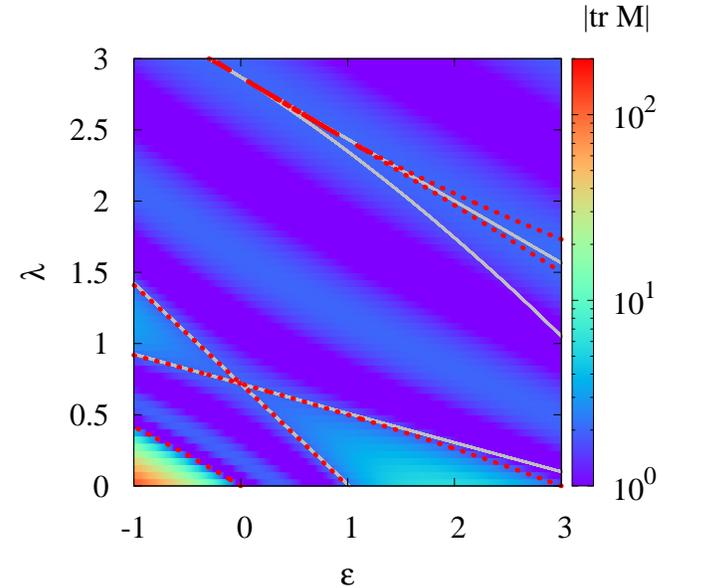}
\caption{
(In)stability boundaries of Lam\'e\rq{}s equation.
The modulus is $k=1/\sqrt{2}$.
Solutions are unstable if $|\tr M| \geq 2$.
The red dotted lines are determined instability boundaries in a basis of Floquet theory which are characterized by $|\tr M| = 2$.
The gray solid lines are given by perturbative calculation (multiple-scale analysis) which is valid when $0 \leq \epsilon < 1$.
}
\label{Lamemap}
\end{figure}

\subsection{Accuracy of multiple-scale analysis}
In this subsection,
we check the accuracy of the multiple-scale analysis for Lam\'e\rq{}s equation for $k=1/\sqrt{2}$.
Figure~\ref{Lamemap} shows the contour map of $|\tr M|$ as a function of $\epsilon,\,\lambda$ where $M$ is a monodromy matrix of Lam\'e\rq{}s equation.
$|\tr M|>2$ means unstable region.
$|\tr M|$ is calculated based on Floquet analysis.
Red dotted lines are contour lines characterized by $|\tr M|=2$ and give \textit{true} instability boundaries.
The boundary lines given by the multiple-scale analysis are plotted with gray solid lines. 

The results of the multiple-scale analysis are in good agreement with the numerically determined boundaries when $\epsilon$ is small enough. 
Let us concentrate on $\epsilon=\pm1, 3$ cases, which are related to Yang-Mills dynamics.
For $\epsilon=1$, the multiple-scale analysis well reproduces the true boundaries.
For $\epsilon=3$, the perturbative approach begins to break down.
According to Fig.~\ref{Lames}, Lam\'e\rq{}s equation with $\epsilon=3$, Eq. \eqref{Lame3}, has only one continuous unstable band while the multiple-scale analysis indicates that there is an unstable band around $\lambda \sim 0$.
For $\epsilon =-1$, the extrapolated boundaries seem to have good accuracy for the sub-dominant band ($0.91 \leq \lambda \leq 1.42$) although $\epsilon <0$ region is out of the validity range of the multiple-scale analysis.

\bibliographystyle{h-physrev5}
\bibliography{Floquetref}
\end{document}